%% file: main.tex
\documentclass[12pt,a4paper]{article}

\usepackage{ifthen} 
\usepackage{rotating}
\usepackage{afterpage}

\newboolean{pdflatex}
\setboolean{pdflatex}{true} 

\newboolean{articletitles}
\setboolean{articletitles}{true} 

\newboolean{uprightparticles}
\setboolean{uprightparticles}{false} 

\newboolean{inbibliography}
\setboolean{inbibliography}{false} 

\input{preamble}

\input{parsedTexTable}

\def\Bctojm {\ensuremath{\Bc\to\jpsi\mup\neum}\xspace}
\def\BctojmX {\ensuremath{\Bc\to\jpsi\mup\neum X}\xspace}
\def\Mjm {\ensuremath{M_{\jpsi\mu}}\xspace}
\def\Mmn {\ensuremath{M_{\mu\nu}}\xspace}
\def\pst {\ensuremath{t_{\mathrm{ps}}}\xspace}
\def\tO  {\ensuremath{t_{0}} \xspace}
\def\pseudot{pseudo-proper time\xspace}
\def\kf {$k$-factor\xspace}
\def\DLL{\ensuremath{\mathrm{DLL}}\xspace}

\def\misbkg{misidentification background\xspace}

\DeclareMathAlphabet{\mathbfit}{OT1}{cmr}{bx}{it}

\begin{document}

\renewcommand{\thefootnote}{\fnsymbol{footnote}}
\setcounter{footnote}{1}

\input{title-LHCb-PAPER}


\renewcommand{\thefootnote}{\arabic{footnote}}
\setcounter{footnote}{0}

\pagebreak


\pagestyle{plain} 
\setcounter{page}{1}
\pagenumbering{arabic}


%

\input{intro}

\input{detector}

\input{selection}

\input{signal}
\input{background}
\input{fit}

\input{syst}

\input{conclusion}

\input{acknowledgements}

\addcontentsline{toc}{section}{References}
\setboolean{inbibliography}{true}
\bibliographystyle{LHCb}
\bibliography{bcsllifetime,main,LHCb-PAPER,LHCb-DP,LHCb-CONF}

\end{document}

%% file: preamble.tex
\usepackage{microtype}
\usepackage{lineno}  
\usepackage{xspace} 

\usepackage{graphicx}  
\usepackage{color}
\usepackage{colortbl}
\graphicspath{{./img/}} 

\usepackage{amsmath} 
\usepackage{amssymb}
\usepackage{amsfonts}
\usepackage{upgreek} 
\usepackage{bm}      

\newcommand*\patchAmsMathEnvironmentForLineno[1]{%
\expandafter\let\csname old#1\expandafter\endcsname\csname #1\endcsname
\expandafter\let\csname oldend#1\expandafter\endcsname\csname
end#1\endcsname
 \renewenvironment{#1}%
   {\linenomath\csname old#1\endcsname}%
   {\csname oldend#1\endcsname\endlinenomath}%
}
\newcommand*\patchBothAmsMathEnvironmentsForLineno[1]{%
  \patchAmsMathEnvironmentForLineno{#1}%
  \patchAmsMathEnvironmentForLineno{#1*}%
}
\AtBeginDocument{%
\patchBothAmsMathEnvironmentsForLineno{equation}%
\patchBothAmsMathEnvironmentsForLineno{align}%
\patchBothAmsMathEnvironmentsForLineno{flalign}%
\patchBothAmsMathEnvironmentsForLineno{alignat}%
\patchBothAmsMathEnvironmentsForLineno{gather}%
\patchBothAmsMathEnvironmentsForLineno{multline}%
}

\usepackage{hyperref}    
\usepackage[all]{hypcap} 

\input{lhcb-symbols-def} 

\usepackage{cite} 
\usepackage{mciteplus}

%% file: lhcb-symbols-def.tex
\def\lhcb {\mbox{LHCb}\xspace}

\ifthenelse{\boolean{uprightparticles}}%
{

 \def\Pmu         {\ensuremath{\upmu}\xspace}                 
 \def\Pnu         {\ensuremath{\upnu}\xspace}                 
                  
 \def\Ppi         {\ensuremath{\uppi}\xspace}

 \def\Ptau        {\ensuremath{\uptau}\xspace}

 \def\Pchi        {\ensuremath{\upchi}\xspace}                 
 \def\Ppsi        {\ensuremath{\uppsi}\xspace}

 \def\PDelta      {\ensuremath{\Delta}\xspace}                 
 \def\PXi      {\ensuremath{\Xi}\xspace}                 
 \def\PLambda      {\ensuremath{\Lambda}\xspace}                 
 \def\PSigma      {\ensuremath{\Sigma}\xspace}                 
 \def\POmega      {\ensuremath{\Omega}\xspace}                 
 \def\PUpsilon      {\ensuremath{\Upsilon}\xspace}                 
 
 \def\PB      {\ensuremath{\mathrm{B}}\xspace}                 
                  
 \def\PD      {\ensuremath{\mathrm{D}}\xspace}

 \def\PJ      {\ensuremath{\mathrm{J}}\xspace}                 
 \def\PK      {\ensuremath{\mathrm{K}}\xspace}

 \def\Pb      {\ensuremath{\mathrm{b}}\xspace}                 
 \def\Pc      {\ensuremath{\mathrm{c}}\xspace}

 \def\Pi      {\ensuremath{\mathrm{i}}\xspace}

 \def\Pp      {\ensuremath{\mathrm{p}}\xspace}

 \def\Ps      {\ensuremath{\mathrm{s}}\xspace}

}
{

 \def\Pmu         {\ensuremath{\mu}\xspace}                 
 \def\Pnu         {\ensuremath{\nu}\xspace}                 
                  
 \def\Ppi         {\ensuremath{\pi}\xspace}

 \def\Ptau        {\ensuremath{\tau}\xspace}

 \def\Pchi        {\ensuremath{\chi}\xspace}                 
 \def\Ppsi        {\ensuremath{\psi}\xspace}                 
                  
 \mathchardef\PDelta="7101
 \mathchardef\PXi="7104
 \mathchardef\PLambda="7103
 \mathchardef\PSigma="7106
 \mathchardef\POmega="710A
 \mathchardef\PUpsilon="7107
                  
 \def\PB      {\ensuremath{B}\xspace}                 
                  
 \def\PD      {\ensuremath{D}\xspace}

 \def\PJ      {\ensuremath{J}\xspace}                 
 \def\PK      {\ensuremath{K}\xspace}

 \def\Pb      {\ensuremath{b}\xspace}                 
 \def\Pc      {\ensuremath{c}\xspace}

 \def\Pi      {\ensuremath{i}\xspace}

 \def\Pp      {\ensuremath{p}\xspace}

 \def\Ps      {\ensuremath{s}\xspace}

}

\def\mup        {\ensuremath{\Pmu^+}\xspace}
\def\mumu       {\ensuremath{\Pmu^+\Pmu^-}\xspace}

\def\taup       {\ensuremath{\Ptau^+}\xspace}

\def\neu        {\ensuremath{\Pnu}\xspace}
\def\neub       {\ensuremath{\overline{\Pnu}}\xspace}
\def\neum       {\ensuremath{\neu_\mu}\xspace}

\def\neut       {\ensuremath{\neu_\tau}\xspace}
\def\neutb      {\ensuremath{\neub_\tau}\xspace}

\def\squark    {\ensuremath{\Ps}\xspace}

\def\cquark    {\ensuremath{\Pc}\xspace}

\def\bquark    {\ensuremath{\Pb}\xspace}
\def\bquarkbar {\ensuremath{\overline \bquark}\xspace}

\def\pion  {\ensuremath{\Ppi}\xspace}

\def\pip   {\ensuremath{\pion^+}\xspace}
\def\pim   {\ensuremath{\pion^-}\xspace}

\def\kaon  {\ensuremath{\PK}\xspace}
  \def\Kbar  {\kern 0.2em\overline{\kern -0.2em \PK}{}\xspace}

\def\Kp    {\ensuremath{\kaon^+}\xspace}
\def\Km    {\ensuremath{\kaon^-}\xspace}

\def\Kstarz  {\ensuremath{\kaon^{*0}}\xspace}

  \def\Dbar    {\kern 0.2em\overline{\kern -0.2em \PD}{}\xspace}
\def\D       {\ensuremath{\PD}\xspace}

\def\Dz      {\ensuremath{\D^0}\xspace}

\def\Dp      {\ensuremath{\D^+}\xspace}

\def\Dstarp  {\ensuremath{\D^{*+}}\xspace}

\def\B       {\ensuremath{\PB}\xspace}
\def\Bbar    {\ensuremath{\kern 0.18em\overline{\kern -0.18em \PB}{}}\xspace}

\def\Bz      {\ensuremath{\B^0}\xspace}

\def\Bu      {\ensuremath{\B^+}\xspace}

\def\Bp      {\ensuremath{\Bu}\xspace}

\def\Bs      {\ensuremath{\B^0_\squark}\xspace}

\def\Bc      {\ensuremath{\B_\cquark^+}\xspace}

\def\jpsi     {\ensuremath{{\PJ\mskip -3mu/\mskip -2mu\Ppsi\mskip 2mu}}\xspace}

  \def\Y#1S{\ensuremath{\PUpsilon{(#1S)}}\xspace}

\def\chic  {\ensuremath{\Pchi_{c}}\xspace}

\def\proton      {\ensuremath{\Pp}\xspace}

\def\Lz {\ensuremath{\PLambda}\xspace}
\def\Lbar {\ensuremath{\kern 0.1em\overline{\kern -0.1em\PLambda}}\xspace}

\def\Lb      {\ensuremath{\Lz^0_\bquark}\xspace}

\def\to                 {\ensuremath{\rightarrow}\xspace}

\def\AT#1     {\ensuremath{A_{\mathrm{T}}^{#1}}\xspace}           

\def\C#1      {\ensuremath{\mathcal{C}_{#1}}\xspace}                       
\def\Cp#1     {\ensuremath{\mathcal{C}_{#1}^{'}}\xspace}                    
\def\Ceff#1   {\ensuremath{\mathcal{C}_{#1}^{\mathrm{(eff)}}}\xspace}        
\def\Cpeff#1  {\ensuremath{\mathcal{C}_{#1}^{'\mathrm{(eff)}}}\xspace}       
\def\Ope#1    {\ensuremath{\mathcal{O}_{#1}}\xspace}                       
\def\Opep#1   {\ensuremath{\mathcal{O}_{#1}^{'}}\xspace}                    

\newcommand{\gev}{\ensuremath{\mathrm{\,Ge\kern -0.1em V}}\xspace}
\newcommand{\tev}{\ifthenelse{\boolean{inbibliography}}{\ensuremath{~T\kern -0.05em eV}\xspace}{\ensuremath{\mathrm{\,Te\kern -0.1em V}}\xspace}}
\newcommand{\mev}{\ensuremath{\mathrm{\,Me\kern -0.1em V}}\xspace}
\newcommand{\kev}{\ensuremath{\mathrm{\,ke\kern -0.1em V}}\xspace}
\newcommand{\ev}{\ensuremath{\mathrm{\,e\kern -0.1em V}}\xspace}
\newcommand{\gevc}{\ensuremath{{\mathrm{\,Ge\kern -0.1em V\!/}c}}\xspace}
\newcommand{\mevc}{\ensuremath{{\mathrm{\,Me\kern -0.1em V\!/}c}}\xspace}
\newcommand{\gevcc}{\ensuremath{{\mathrm{\,Ge\kern -0.1em V\!/}c^2}}\xspace}
\newcommand{\gevgevcccc}{\ensuremath{{\mathrm{\,Ge\kern -0.1em V^2\!/}c^4}}\xspace}
\newcommand{\mevcc}{\ensuremath{{\mathrm{\,Me\kern -0.1em V\!/}c^2}}\xspace}

\def\mum  {\ensuremath{{\,\upmu\rm m}}\xspace}

\def\invfb   {\ensuremath{\mbox{\,fb}^{-1}}\xspace}

\newcommand{\stat}{\ensuremath{\mathrm{\,(stat)}}\xspace}
\newcommand{\syst}{\ensuremath{\mathrm{\,(syst)}}\xspace}

\newcommand{\chisq}{\ensuremath{\chi^2}\xspace}

\def\gsim{{~\raise.15em\hbox{$>$}\kern-.85em
          \lower.35em\hbox{$\sim$}~}\xspace}
\def\lsim{{~\raise.15em\hbox{$<$}\kern-.85em
          \lower.35em\hbox{$\sim$}~}\xspace}

\def\PDF {PDF\xspace}

\def\pt         {\mbox{$p_{\rm T}$}\xspace}

\def\evtgen     {\mbox{\textsc{EvtGen}}\xspace}

\def\geant      {\mbox{\textsc{Geant4}}\xspace}

\def\photos     {\mbox{\textsc{Photos}}\xspace}

\def\pythia     {\mbox{\textsc{Pythia}}\xspace}

\def\tell1  {TELL1\xspace}
\def\ukl1   {UKL1\xspace}



%% file: parsedTexTable.tex
\newcommand{\CENTRALVALUE}{508.7}

\newcommand{\offsetCENTRALVALUENP}{\ensuremath{+6.4}}
\newcommand{\systCENTRALVALUENP}{6.4}

\newcommand{\offsetFREEMISIDNORM}{\ensuremath{+0.8}}
\newcommand{\systFREEMISIDNORM}{0.8}

\newcommand{\systBBBARNORMINCREASED}{3.4}

\newcommand{\systFAKEJPSINORM}{0.4}

\newcommand{\offsetBBBARSINGLEEXPONENTIAL}{\ensuremath{ -7.3}}
\newcommand{\systBBBARSINGLEEXPONENTIAL}{7.3}

\newcommand{\systCUTWRONGPV}{1.8}

\newcommand{\offsetUSEEBERT}{\ensuremath{+2.0}}

\newcommand{\offsetUSEISGWTWO}{\ensuremath{ -1.5}}

\newcommand{\systNOMOMENTUMSCALECALIBRATION}{0.2}

\newcommand{\systFAKEJPSINORIGHTSB}{2.3}

\newcommand{\offsetMISIDDEFORMRB}{\ensuremath{ -1.2}}
\newcommand{\systMISIDDEFORMRB}{1.2}

\newcommand{\offsetSIGNALMODELFOURGRES}{\ensuremath{ -1.3}}
\newcommand{\systSIGNALMODELFOURGRES}{1.3}

\newcommand{\doSystTable}{
\begin{tabular}{lr}
\hline
Source & Assigned systematic [fs] \\
\hline
\Bc production  model& 1.0 \\

\Bc decay model& 5.0 \\

Signal resolution model&\systSIGNALMODELFOURGRES\\
Prompt background model&\systCENTRALVALUENP\\
Fake \jpsi background yield&\systFAKEJPSINORM\\
Fake \jpsi background shape&\systFAKEJPSINORIGHTSB\\
Combinatorial background yield&\systBBBARNORMINCREASED\\
Combinatorial background shape&\systBBBARSINGLEEXPONENTIAL\\
Misidentification background yield&\systFREEMISIDNORM\\
Misidentification background shape&\systMISIDDEFORMRB\\
Length scale calibration& 1.3 \\

Momentum scale calibration&\systNOMOMENTUMSCALECALIBRATION\\
Efficiency function& 2.6 \\

Incorrect association to PV&\systCUTWRONGPV\\
Multiple candidates& 1.0 \\

Fit validation& 0.5 \\

\hline Quadratic sum  &  \SYSTERROR \\\hline

\end{tabular} 
}

\newcommand{\CENTRALVALUEshort}{509}
\newcommand{\STATERROR}{7.7}

\newcommand{\SYSTERROR}{12.4}
\newcommand{\STATERRORshort}{8}
\newcommand{\SYSTERRORshort}{12}

%% file: title-LHCb-PAPER.tex

\begin{titlepage}
\pagenumbering{roman}

\vspace*{-1.5cm}
\centerline{\large EUROPEAN ORGANIZATION FOR NUCLEAR RESEARCH (CERN)}
\vspace*{1.5cm}
\hspace*{-0.5cm}
\begin{tabular*}{\linewidth}{lc@{\extracolsep{\fill}}r}
\ifthenelse{\boolean{pdflatex}}
{\vspace*{-2.7cm}\mbox{\!\!\!\includegraphics[width=.14\textwidth]{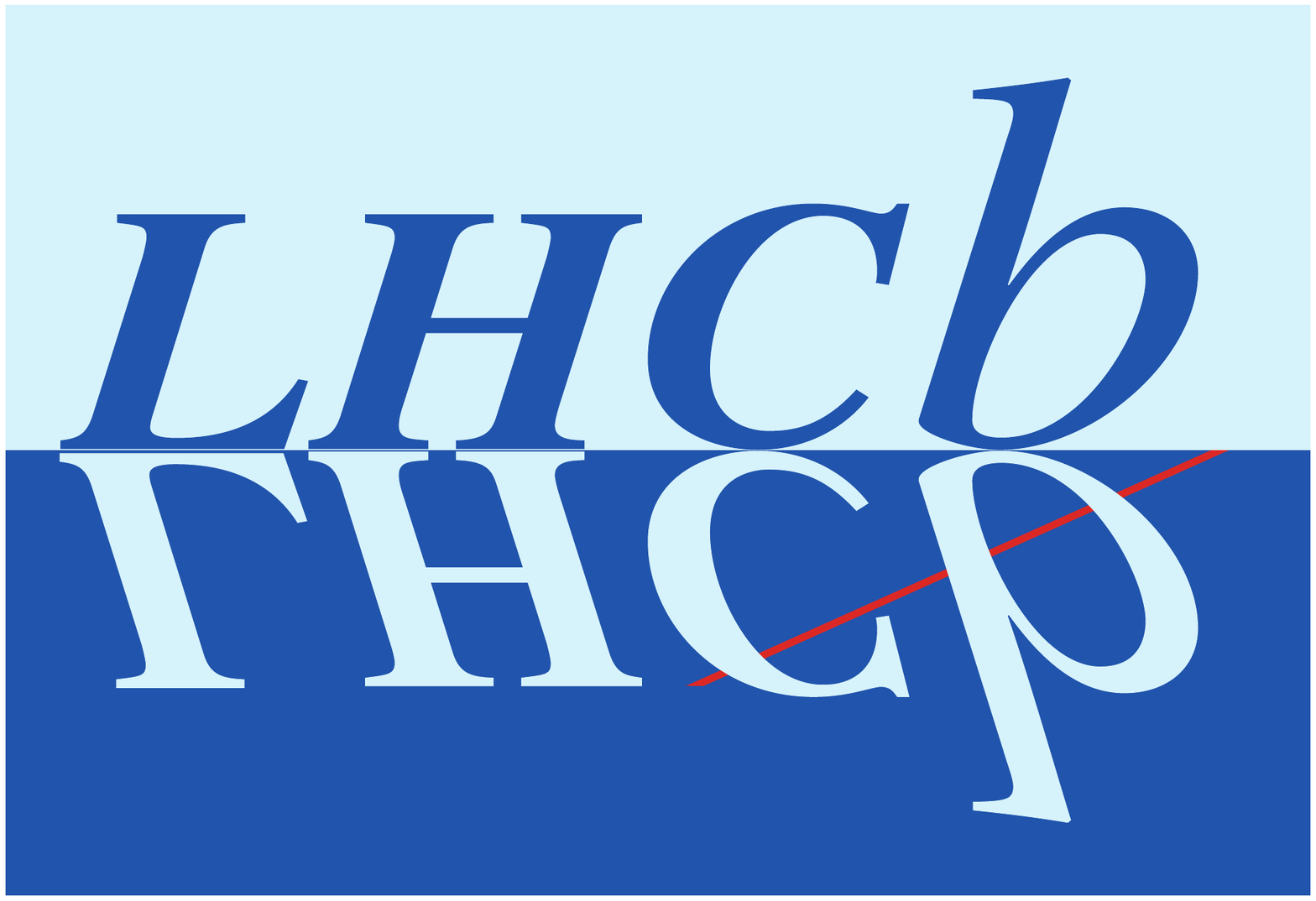}} & &}%
{\vspace*{-1.2cm}\mbox{\!\!\!\includegraphics[width=.12\textwidth]{img/lhcb-logo.eps}} & &}%
\\
 & & CERN-PH-EP-2014-008 \\  
 & & LHCb-PAPER-2013-063 \\  
 & & 28 January 2014  \\ 
\end{tabular*}

\vspace*{4.0cm}

{\bf\boldmath\huge
\begin{center}
Measurement of the \Bc meson lifetime using \BctojmX decays
\end{center}
}

\vspace*{2.0cm}

\begin{center}
The LHCb collaboration\footnote{Authors are listed on the following pages.}
\end{center}

\vspace{\fill}

\begin{abstract}
  \noindent
  The lifetime of the \Bc meson is measured using semileptonic decays having 
  a \jpsi meson and a muon in the final state.
The data, corresponding to an integrated luminosity of $2\invfb$, are collected by the 
\lhcb detector in $pp$ collisions at a centre-of-mass energy of $8\tev$.  The measured
lifetime is
  $$\tau = \CENTRALVALUEshort \pm \STATERRORshort  \pm \SYSTERRORshort   \mathrm{~fs}, $$
  where the first uncertainty is statistical and the second is systematic. 
\end{abstract}

\vspace*{2.0cm}

\begin{center}
  Submitted to Eur.~Phys.~J.~C 
\end{center}

\vspace{\fill}

{\footnotesize 
\centerline{\copyright~CERN on behalf of the \lhcb collaboration, license \href{http://creativecommons.org/licenses/by/3.0/}{CC-BY-3.0}.}}
\vspace*{2mm}

\vspace{\fill}

\end{titlepage}


\newpage
\setcounter{page}{2}
\mbox{~}
\newpage

\input{LHCb_HD_authorlist_2013-11-22.tex}

\cleardoublepage

%% file: LHCb_HD_authorlist_2013-11-22.tex
\centerline{\large\bf LHCb collaboration}
\begin{flushleft}
\small
R.~Aaij$^{40}$, 
B.~Adeva$^{36}$, 
M.~Adinolfi$^{45}$, 
A.~Affolder$^{51}$, 
Z.~Ajaltouni$^{5}$, 
J.~Albrecht$^{9}$, 
F.~Alessio$^{37}$, 
M.~Alexander$^{50}$, 
S.~Ali$^{40}$, 
G.~Alkhazov$^{29}$
% ...

{\footnotesize \it
% ...
$ ^{5}$Clermont Universit\'{e}, Universit\'{e} Blaise Pascal, CNRS/IN2P3, LPC, Clermont-Ferrand, France\\
% ...
$ ^{9}$Fakult\"{a}t Physik, Technische Universit\"{a}t Dortmund, Dortmund, Germany\\
% ...
$ ^{29}$Petersburg Nuclear Physics Institute (PNPI), Gatchina, Russia\\
% ...
$ ^{36}$Universidad de Santiago de Compostela, Santiago de Compostela, Spain\\
$ ^{37}$European Organization for Nuclear Research (CERN), Geneva, Switzerland\\
% ...
$ ^{40}$Nikhef National Institute for Subatomic Physics, Amsterdam, The Netherlands\\
% ...
$ ^{45}$H.H. Wills Physics Laboratory, University of Bristol, Bristol, United Kingdom\\
% ...
$ ^{50}$School of Physics and Astronomy, University of Glasgow, Glasgow, United Kingdom\\
$ ^{51}$Oliver Lodge Laboratory, University of Liverpool, Liverpool, United Kingdom\\
% ...
}
\end{flushleft}

%% file: intro.tex

\section{Introduction}
\label{sec:Introduction}

The \Bc meson, formed of a \bquarkbar and a \cquark quark,  is an excellent laboratory
to study QCD and weak interactions.\footnote{The inclusion of charge conjugate states is always implied throughout 
this paper.}   The \Bc meson bound-state dynamics
can be treated in a non-relativistic expansion by QCD-inspired effective  
models that successfully describe the spectroscopy of quarkonia. However,
\Bc production and decay dynamics have some distinctive features, since this meson is 
the only observed open-flavour state formed  by two heavy quarks. 
The decay proceeds through the weak interaction, and about 70\% of the width
is expected to be due to the CKM favoured $c \to s$ transition~\cite{Gouz:2002kk}. This decay process,
challenging to detect, has recently been observed  in the $\Bc\to \Bs\pip$ mode
by the \lhcb collaboration\cite{LHCb-PAPER-2013-044}.
The $b \to c$ transition  offers an easier experimental signature,
having a substantial probability to produce a \jpsi meson.
Indeed, the \Bc meson was discovered by the CDF
collaboration~\cite{Abe:1998wi} through the observation of the
 $\Bc\to\jpsi\ell^+\nu_{\ell}X ~(\ell = \mu, e)$ semileptonic
decays, where $X$ denotes any possible additional
particles in the final state.

The precise measurement of the \Bc lifetime provides an essential test of the
theoretical models describing its dynamics.
Computations based on various
frameworks~\cite{Beneke:1996xe,Anisimov:1998uk,Kiselev:2000pp,Chang:2000ac,Gouz:2002kk}
predict values ranging from 300 to 700 fs. 
The world average value of the \Bc lifetime reported by the PDG in
2013~\cite{PDG2012} is 452 $\pm$ 33 fs. This was obtained from measurements
performed at the Tevatron, using semileptonic decays~\cite{Abe:1998wi,Abulencia:2006zu,Abazov:2008rba} and 
the rarer  $B_c^+ \to \jpsi \pi^+$ decay~\cite{Aaltonen:2012yb}.

The unprecedented \Bc production rate achieved at the LHC 
has thus far been used to measure many \Bc decay properties,
 with several new decay modes
observed by \lhcb~\cite{LHCb-PAPER-2011-044, LHCb-PAPER-2012-054, LHCb-PAPER-2013-010,
  LHCb-PAPER-2013-021, LHCb-PAPER-2013-047, LHCb-PAPER-2013-044}.  The current knowledge of the lifetime is one of the largest systematic uncertainties in the relative branching
fraction measurements, also affecting the determination of
the production cross-section~\cite{LHCb-PAPER-2012-028}.
This paper reports a measurement of the \Bc lifetime using the
semileptonic decays $\Bc\to\jpsi\mup\neum X$ with $\jpsi\to\mumu$.

%% file: detector.tex
\section{Detector and data sample}
\label{sec:Detector}

The \lhcb detector~\cite{Alves:2008zz} is a single-arm forward
spectrometer covering the \mbox{pseudorapidity} range $2<\eta <5$,
designed for the study of particles containing \bquark or \cquark
quarks. The detector includes a high-precision tracking system
consisting of a silicon-strip vertex detector surrounding the $pp$
interaction region, a large-area silicon-strip detector located
upstream of a dipole magnet with a bending power of about
$4{\rm\,Tm}$, and three stations of silicon-strip detectors and straw
drift tubes placed downstream.
The combined tracking system provides a momentum measurement with
relative uncertainty that varies from 0.4\% at 5\gevc to 0.6\% at 100\gevc,
and impact parameter resolution of 20\mum for
tracks with large transverse momentum. Different types of charged hadrons are distinguished by information
from two ring-imaging Cherenkov detectors~\cite{LHCb-DP-2012-003}. Photon, electron and
hadron candidates are identified by a calorimeter system consisting of
scintillating-pad and preshower detectors, an electromagnetic
calorimeter and a hadronic calorimeter. Muons are identified by a
system composed of alternating layers of iron and multiwire
proportional chambers~\cite{LHCb-DP-2012-002}.
The trigger~\cite{LHCb-DP-2012-004} consists of a
hardware stage, based on information from the calorimeter and muon
systems, followed by a software stage, which applies a full event
reconstruction.

The analysis is performed on a data sample of $pp$ collisions at a
centre-of-mass energy of 8 TeV,  collected during 2012
and corresponding to an integrated luminosity of 2\,fb$^{-1}$.
Simulated event samples are generated for the signal decays and 
the decay modes contributing to the background.
In the simulation, $pp$ collisions are generated using
\pythia~\cite{Sjostrand:2006za}  with a specific \lhcb
configuration~\cite{LHCb-PROC-2010-056}.  The production of 
\Bc mesons, which is not adequately simulated in \pythia, is performed 
by the dedicated generator \mbox{\textsc{Bcvegpy}}\cite{Chang:2003cq} 
using a \Bc mass of 6276 \mevcc and a lifetime of 450 fs. 
Several dynamical models are
used to simulate \Bctojm decays,  as discussed in Sec.~\ref{sec:Signal}.
Decays of hadronic particles
are described by \evtgen~\cite{Lange:2001uf}, in which final state
radiation is generated using \photos~\cite{Golonka:2005pn}. 
The interaction of the generated particles with the detector and its
response are implemented using the \geant
toolkit~\cite{Allison:2006ve, *Agostinelli:2002hh} as described in
Ref.~\cite{LHCb-PROC-2011-006}.

%% file: selection.tex
\section {Analysis strategy and event selection} 
\label{sec:Selection}

Candidate signal decays are obtained from combinations of 
a dimuon compatible with a \jpsi decay and an additional candidate
muon track, denoted as a {\it bachelor} muon in the following, originating from
a common vertex.

Since the expected signal  yield is about $10^4$
candidates over a moderate background, the  event selection and analysis are
driven by the need to minimise systematic uncertainties. Selection variables 
that bias the \Bc candidate decay time distribution are avoided,
and the selection is designed not only  
to  suppress the background contributions, but also to allow their
modelling using data.
Background candidates with decay time and $\jpsi\mu$ mass values comparable to
the signal decays are
mainly expected from \bquark-hadron decays to a \jpsi meson and a
hadron that is misidentified as a muon.
This misidentification
background is modelled using data  in which \Bc candidates are
selected without any bias related to the identification of the bachelor muon.
The candidate events are required to pass a trigger decision based
solely on the information from the \jpsi\to\mumu candidate. To pass the hardware trigger,
one or both tracks from the \jpsi decay must be identified as muons.
In the first case, the muon is required to have a transverse momentum, \pt, greater than 
$1.48\gevc$,
while in the second case, the product of the two \pt values must be larger than $1.68\gev^2/c^2$.
The software trigger selects dimuon candidates
consistent with the decay of a \jpsi meson by applying loose criteria
on the dimuon mass, vertex quality and muon identification, and requires $\pt>2\gevc$.

An offline selection applies
further kinematic criteria to enhance the signal purity.
Requirements on the minimum \pt are applied to the two
\jpsi decay products ($1.4\gevc$), the \jpsi candidate ($2\gevc$),
the bachelor muon ($2.5\gevc$) and the $\jpsi\mu$ combination ($6\gevc$).
The momentum of the bachelor muon must be between 13 and 150 \gevc.
The \jpsi candidate mass  is required to be between 3.066 and 3.131
\gevcc, a range corresponding to about four times the mass resolution.
Two sideband mass regions, 3.005--3.036 and 3.156--3.190 \gevcc,
are used to evaluate the background from track pairs misidentified as \jpsi 
candidates.  
The three muons  are required to originate from a common vertex, with a 
\chisq per degree of freedom from the fit smaller than 3.0. This restrictive requirement
suppresses combinatorial background from random associations of real
\jpsi and muon candidates not originating from the same vertex.
The $\jpsi\mu$  mass, \Mjm, is reconstructed from a kinematic fit
constraining the \jpsi mass to its known value\cite{PDG2012}, and is
required to be between 3.5 and 6.25 \gevcc.

Particle identification is based on the information from the Cherenkov, 
calorimeter and muon detectors, combined into likelihood functions.
The selection is based on the logarithm of the likelihood ratio, $\DLL_{P/P^{'}}$, for two given
 charged-particle hypotheses $P$ and $P^{'}$ among $\mu, \pi, K$ and $p$.
The requirement $\DLL_{\mu/\pi}>1$ is applied on the two muon tracks forming the \jpsi candidate.
Dedicated, more restrictive identification requirements are applied to the bachelor
muon candidate, including  the criterion that the track
is matched with muon detector hits in all stations
downstream of the calorimeters. A track fit based on a Kalman filter~\cite{kalman} is performed using such 
hits, and the resulting  $\chisq$ per degree of freedom must be lower than 1.5.
Vetoes against the pion ($\DLL_{\mu/\pi}>3$), kaon
($\DLL_{K/\pi}<8$) and proton ($\DLL_{p/\pi}<20$) hypotheses are also applied.
To avoid cases in which two candidate tracks are
reconstructed from the same muon, the bachelor candidate is required not to
share any hits in the muon detectors, and share less than 20\% of
hits in the tracking stations, with either of the two other muon candidates in
the decay.   
Studies using simulated samples indicate that, after these requirements, the
misidentified candidates are dominated
by kaons and pions decaying in flight. Decays
occurring in the tracker region are reduced by requiring a good match
between the track segments reconstructed upstream and downstream of the
magnet ($\chisq<15.0$ with five degrees of freedom).

The  selected sample consists of 29\,756 candidates. Among the
selected events, 0.6\% have multiple candidates which, in most cases,
are formed by the same three tracks where the muons with the same
charge are exchanged. All candidates are 
retained, and this effect is considered as a
potential source of systematic uncertainty.

To study the decay time distribution,  a \pseudot{} is determined for each candidate, defined as
\begin{equation} \label{eq:pst}
  \pst = 
\mathbfit{p} \cdot (\mathbfit{v}-\mathbfit{x})
                \frac{M_{3\mu}}{|\mathbfit{p}|^2},
\end{equation}
where  $\mathbfit p$ is the three-momentum of the  $\jpsi\mu$ system in the laboratory frame,
and $\mathbfit{v}$ and $\mathbfit{x}$ are the  measured positions of the
\Bc decay and production vertices, respectively. The primary $pp$ interaction vertex
(PV) associated with the production of each \Bc candidate is chosen as
the one yielding the smallest  difference in \chisq when fitted  with and
without the \Bc candidate. The position obtained from the latter
fit is used. The three-$\mu$ mass $M_{3\mu}$ used in Eq.~\ref{eq:pst} is computed
without the constraint on the \jpsi mass, to reduce the potential  bias
from momentum scale miscalibration, which
approximately cancels in the $M_{3\mu}/|\mathbfit{p}|$ ratio.

The \Bc lifetime is determined using the variables \pst and \Mjm.
To infer the \Bc decay time from the \pseudot{}, a statistical
correction based on simulation, commonly referred to as the \kf method, is adopted.
There, the average effect of the  momentum of the unreconstructed decay products
on the determination of the \Bc decay time is computed as a function of \Mjm. The \Bc momentum can also be
reconstructed for each decay, up to a two-fold ambiguity, using the
measured flight direction of the \Bc meson and the knowledge of its mass. However, due to the short
\Bc lifetime, the achievable resolution is poor and strongly dependent
on the decay time. Therefore, this partial reconstruction is
not used in the lifetime determination to avoid potentially large
biases, but exploited to study systematic 
uncertainties arising from the assumed kinematic model of the signal.

The background contributions are also modelled in
the  (\pst, \Mjm) plane. Models are obtained 
from data whenever possible, with the notable exception of
combinatorial background, whose contribution is inferred from large simulated
samples of inclusive \bquark-hadron decays containing a \jpsi meson
in the final state.

%% file: signal.tex
\section{Signal model}
\label{sec:Signal}

The expected (\pst, \Mjm) distribution for the signal decays depends on the simulation 
of the dynamics for the \Bctojm decay and of the contributions 
from decay modes with additional particles in the final state ({\it feed-down} modes).

\begin{figure}[tb]
	\centering
	\includegraphics[width=0.7\textwidth,trim=0 2mm 0 0,clip]{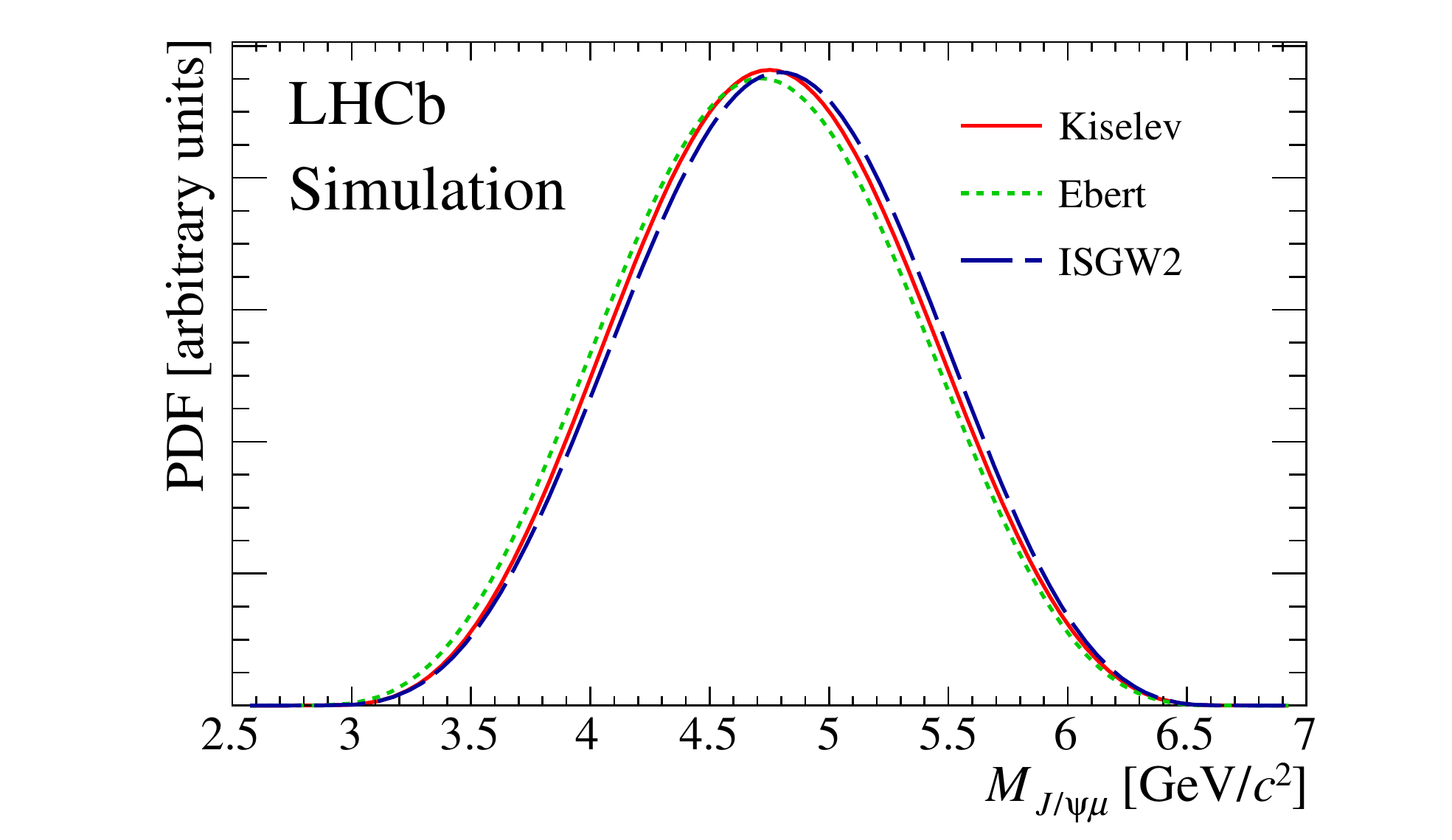}
	\caption{\small Distributions of \Mjm, without simulation of detector response, 
          for the Kiselev (red solid line),  Ebert (green short-dashed line), and  
          ISGW2 (black long-dashed line) models. 
        }
	\label{fig:sigGenLevelMass}
\end{figure}

For the \Bctojm decay, three different decay models are
implemented in the simulation, referred to as  
Kiselev~\cite{Kiselev:1993ea,Kiselev:1999sc,Kiselev:2002vz}, Ebert\cite{Ebert:2003cn}
and ISGW2~\cite{PhysRevD.52.2783}. The Kiselev model
is adopted as the baseline and used to simulate more than 20 million events
with the three muons in the nominal detector acceptance.
Smaller samples generated with the alternative models are used for systematic studies. Figure~\ref{fig:sigGenLevelMass}
compares the probability density function (\PDF) for \Mjm  predicted by  the three models, which exhibit only small differences with each other.

The simulation is used to predict  the average ratio between 
the  measured \pseudot and the simulated true
\Bc decay time $t^*$. This correction term can be factorised as 
\begin{equation}
  \label{eq:kfactor}
  k' \equiv \frac{\pst}{t^*} = \frac{\pst}{\pst^*} \times   \frac{\pst^*}{t^*} 
  \equiv \frac{\pst}{\pst^*} \times k,
\end{equation}
where $\pst^*$ is the simulated true value of the \pseudot defined in
Eq.~\ref{eq:pst}. 
The $\pst/\pst^*$ term accounts for imperfections in
the experimental reconstruction, while the 
$k\equiv \pst^*/t^*$ factor includes only the kinematic effects from unobserved 
particles in the final state. It is found that the kinematic term dominates
the average deviation from unity and the r.m.s. width of the $k'$
variable. The \kf distribution is empirically modelled from simulated events in bins of \Mjm.

The resolution function describing  $\Delta t\equiv\pst-\pst^*$ is parametrised as the sum
of three Gaussian functions with a common mean $t_0$ and different widths
 \begin{equation}\label{eq:resolutionmodel}
  G(\Delta t) = \sum_{i=1}^3 g_i ~\frac{1}{\sigma_i \sqrt{2\pi}} \exp\left(-\frac{(\Delta t-t_0)^2}{2\sigma_i^2}\right).
\end{equation}
The parameters $g_i$, $t_0$ and $\sigma_i$ are determined from fits to
the simulated events. A small bias $t_0=-1.9 \pm 0.2$~fs is found, 
and the core Gaussian term has parameters $g_1=0.74$, $\sigma_1=27$~fs.
The other two Gaussian functions have parameters $g_2=0.24$, $\sigma_2=54$~fs, $g_3=0.02$, and $\sigma_3=260$~fs.
These parameters are assumed not to
depend on the decay time itself as indicated by the simulation.
A fourth Gaussian term with the same mean and large width is added when 
performing the fit  
to simulated data to describe the small fraction of events
having an incorrectly associated primary vertex. This is not included in the signal
model because these events are considered as a
background source, which is constrained from data
by exploiting the negative tail of the \pst distribution.

The model for the \PDF  of $\pst=kt^*
+ \Delta t$ is obtained for each \Mjm bin $m$ by convoluting
the exponential $t^*$ distribution with the \kf distribution
$h^{m}(k)$ and the  resolution function, resulting in
\begin{align}\label{eq:sigPstmodel}
  f^m(\pst) 
  =& \sum_{i=1}^3 g_i \int_{-\infty}^{+\infty}\mathrm dk\, h^{m}(k) 
  \frac{1}{2k\tau} 
    \exp\left(\frac{\sigma_i^2}{2k^2 \tau^2} - \frac{(\pst - \tO)}{k\tau}\right)
  \mathrm{erfc}\left(\frac{\sigma_i}{k\tau\sqrt{2}} - \frac{(\pst - \tO)}{\sigma_i\sqrt{2}}\right),
\end{align}
where $\tau$ is the \Bc lifetime and erfc is the complementary error function. 

\begin{figure}[tb]
  \centering
  \begin{minipage}{0.54\linewidth}
    \includegraphics[width=\textwidth]{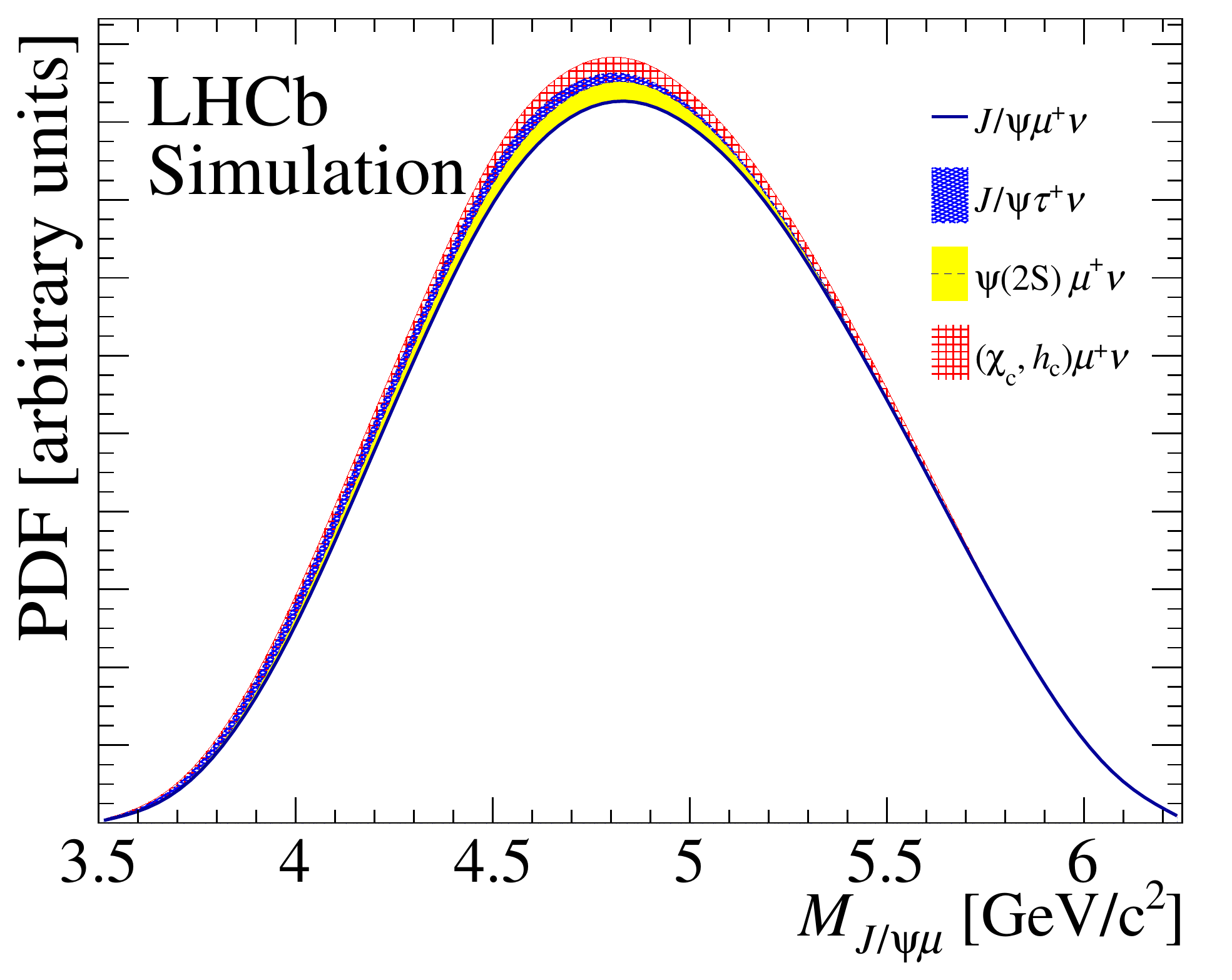}
    \put(-215,140){$\mathrm{a})$}
  \end{minipage}
  \begin{minipage}{0.45\linewidth}
  \includegraphics[width=\textwidth]{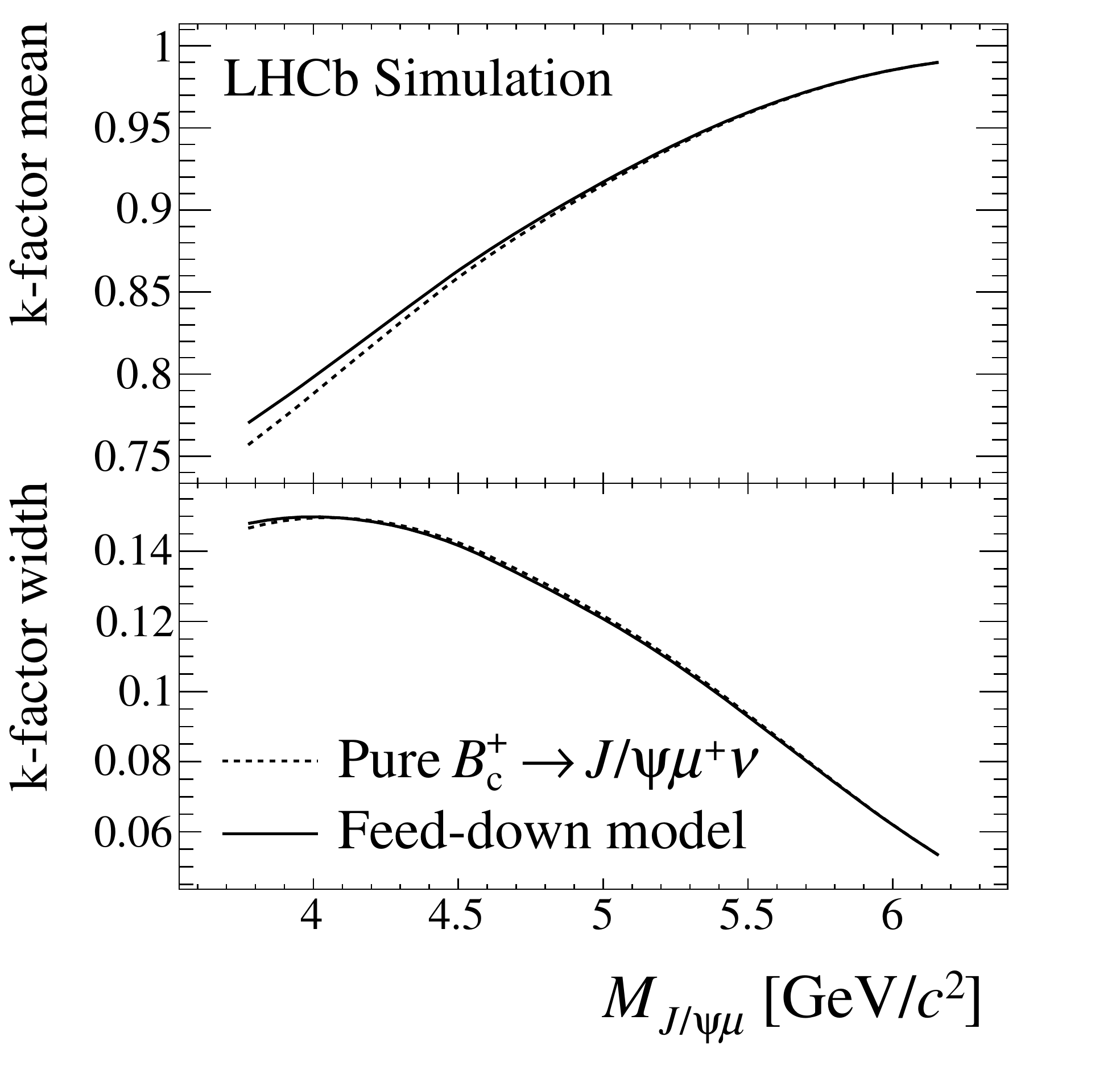}
    \put(-163,165){$\mathrm{b})$}
  \end{minipage}
  \caption{\small  \label{fig:feeddowns}
    Corrections to the (a) \Mjm and (b) \kf model due to the contribution of
    feed-down modes after the selection. The contribution to the \Mjm distribution from \Bctojm decays alone
    is shown by the black solid curve, while the
    inclusion of each modelled feed-down contribution is shown by the shaded areas
    according to the legend.
    The mean and r.m.s. width of the \kf
    distribution are shown as a function of \Mjm before (solid line) and after
    (dashed line) the inclusion of feed-down modes. 
  } 
\end{figure}
The signal model must also consider feed-down modes. Their contribution 
is expected to bias the measured lifetime by modifying
the \Mjm distribution towards lower values and, to a lesser extent, by
affecting the \kf distribution.  
The modes  explicitly included in the model are semileptonic \Bc decays to
the higher charmonia
states $\psi(2S)$, $\chi_{cJ} ~(J=0,1,2)$ and $h_{c}$, subsequently
decaying to  a $\jpsi X$ final state, and the $B_c^+ \to \jpsi \taup \neut$ decay followed by $\taup \to \mup \neum \neutb$.
Theoretical calculations give the decay widths of these decay chains
relative to \Bctojm to be 3.0\% for the $\psi(2S)$ mode~\cite{Kiselev:2002vz},   
3.3\% for the sum of \chic and $h_c$
contributions~\cite{Wang:2011jt,Ebert:2010zu,Kuang:1988bz,Andreotti:2005vu},
and 
4.4\% for $\jpsi\taup\neut$ decays~\cite{Gouz:2002kk}. The first two of these are subject to large uncertainties, which are considered 
in the systematic uncertainty. 
The contribution of the feed-down modes after the selection 
is found to be small, as shown in Fig.~\ref{fig:feeddowns}.

Other possible feed-down contributions are the abundant
$\Bc\to\Bs\mup\neum$ decay mode, followed by the $\Bs\to\jpsi X$ decay, 
and decays to $\jpsi \Dp_{(s)}$ final states followed by the semileptonic 
decay of the charmed meson. These channels are studied using
simulated events and found to be negligible, mainly because of the softer
\pt spectrum of the bachelor muon, and the long-lived intermediate
particle causing the reconstructed three-muon vertex to be of poor quality.

%% file: background.tex
\section{Background model}
\label{sec:Background}
The main background to decays of long lived particles to three muons is expected
to be due to hadrons misidentified as muons and combined with a 
\jpsi meson from the same vertex, hereafter referred to as {\it\misbkg}.
Other sources of background, with a correctly identified bachelor
muon, are either due to false \jpsi candidates  ({\it fake \jpsi
background} in the following), or associations of a genuine \jpsi
meson and
a real bachelor muon not originating from \Bc decays. In the latter case,
the two particles can both be produced at the PV ({\it prompt
background}), produced at different vertices and randomly
associated ({\it combinatorial background}), or produced at the same
detached vertex ({\it $B\to3\mu$ background}).
The yield and \PDF of each contribution is modelled from data,
with the exception of the last two categories, where simulation is used.

The \misbkg can be accurately predicted from data as no
identification requirements are imposed on the bachelor muon by the trigger.
By also removing such requirements from the
offline selection, a {\it \jpsi-track} sample consisting of  $5.5\times 10^6$ candidates,
dominated by \jpsi-hadron combinations, is obtained.
The \misbkg is modelled by weighting each candidate in this sample by 
$W$, the probability to misidentify a hadron as a bachelor muon candidate.
This is defined as the average over  hadron species $h$  of
the misidentification probability $W_h$  for the given species,
each being weighted by the probability $P_h$ for the track to be a hadron $h$
\begin{equation}\label{eq:misidW}
	W =  \sum_{h = K, \pi, p} P_h(\eta, p_{h}, I)
        W_h(\eta, p_{h}, N_t),
\end{equation}
where $h$ can be a kaon, a pion or a proton. 
The quantities $P_h$ and $W_h$ are measured using calibration samples,
as functions of the most relevant variables on which they depend.
For $P_h$, these are the track momentum $p_{h}$, its pseudorapidiy $\eta$,
and the impact parameter $I$ with respect to the PV. The dependence on
 $I$ arises because particles produced at the
collision vertex will prevail around the PV position, while \bquark-hadron decays 
dominate the events with a sizeable $I$ value.

For $W_h$, the variables are $p_{h}$, $\eta$ and the number of tracks in
the event $N_t$, since the  particle identification performance,
notably for the Cherenkov detectors, is affected by the density of
hits.
The contribution from cases where the bachelor track in the \jpsi-track sample 
is a lepton is  neglected, since its effect on the predicted background yield
is small compared to the final statistical and systematic uncertainties.
Calibration samples consist of selected $\Dstarp \to \pip \Dz
(\Km\pip)$ decays for kaons and pions, and $\Lz\to\proton\pim$ 
decays for protons. 
The residual background to these selections, at the level of a few per cent, is subtracted using
events in the sidebands of the $D$ or $\Lz$ mass distributions. 

\begin{figure}[bt]
  \centering
  \includegraphics[width=.8\textwidth]{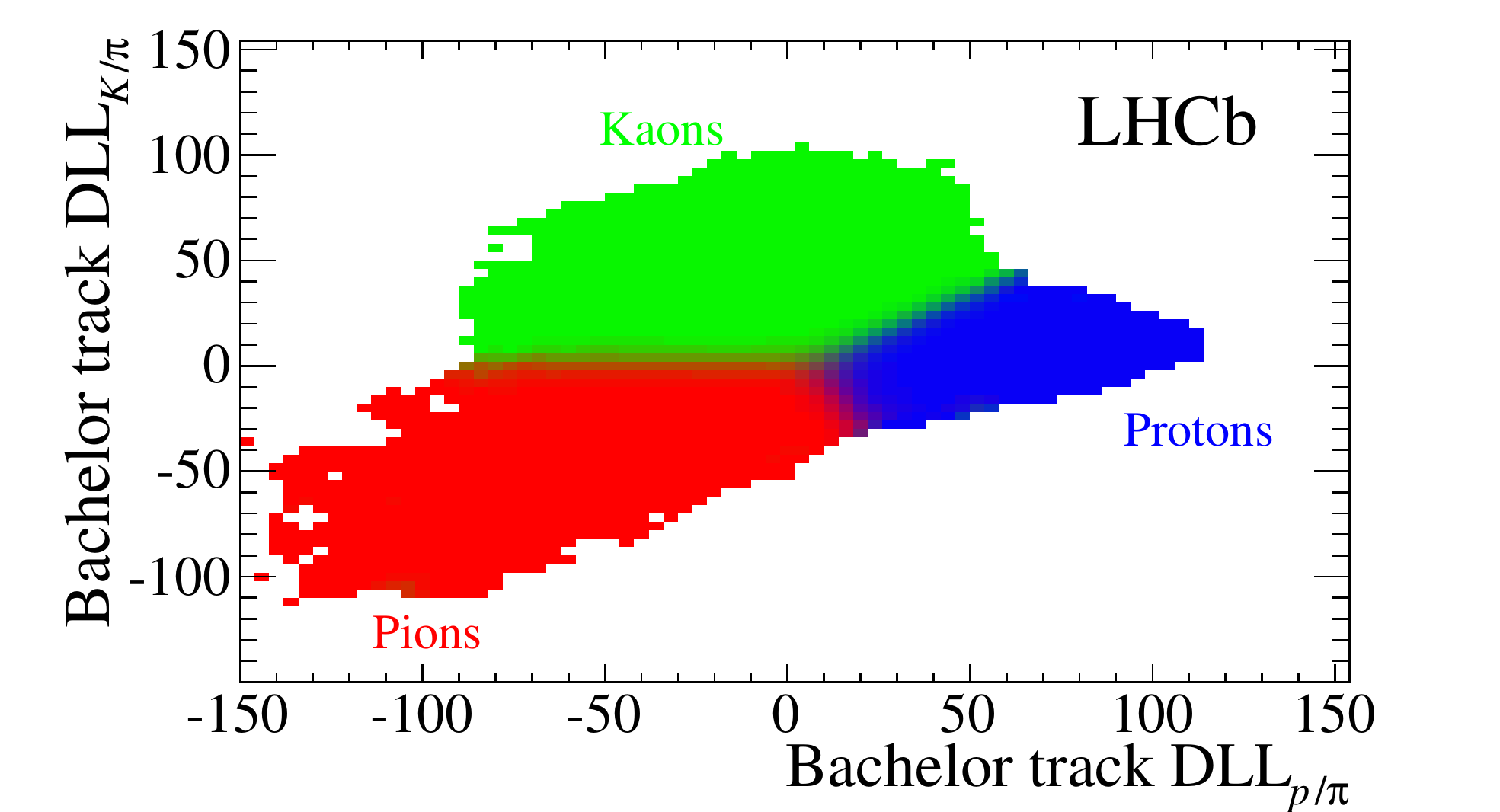}
  \caption{\small \label{fig:planeLikelihoods}
    Result of a fit in the ($\DLL_{p/\pi}, \DLL_{K/\pi}$) plane to determine
    the fractions of hadron species in the \jpsi-track sample.
    The colour of each bin is built as
    a combination of red, green and blue proportional to the fitted fractions
    of pions, kaons and protons, respectively.
  }
  \includegraphics[width=.8\textwidth]{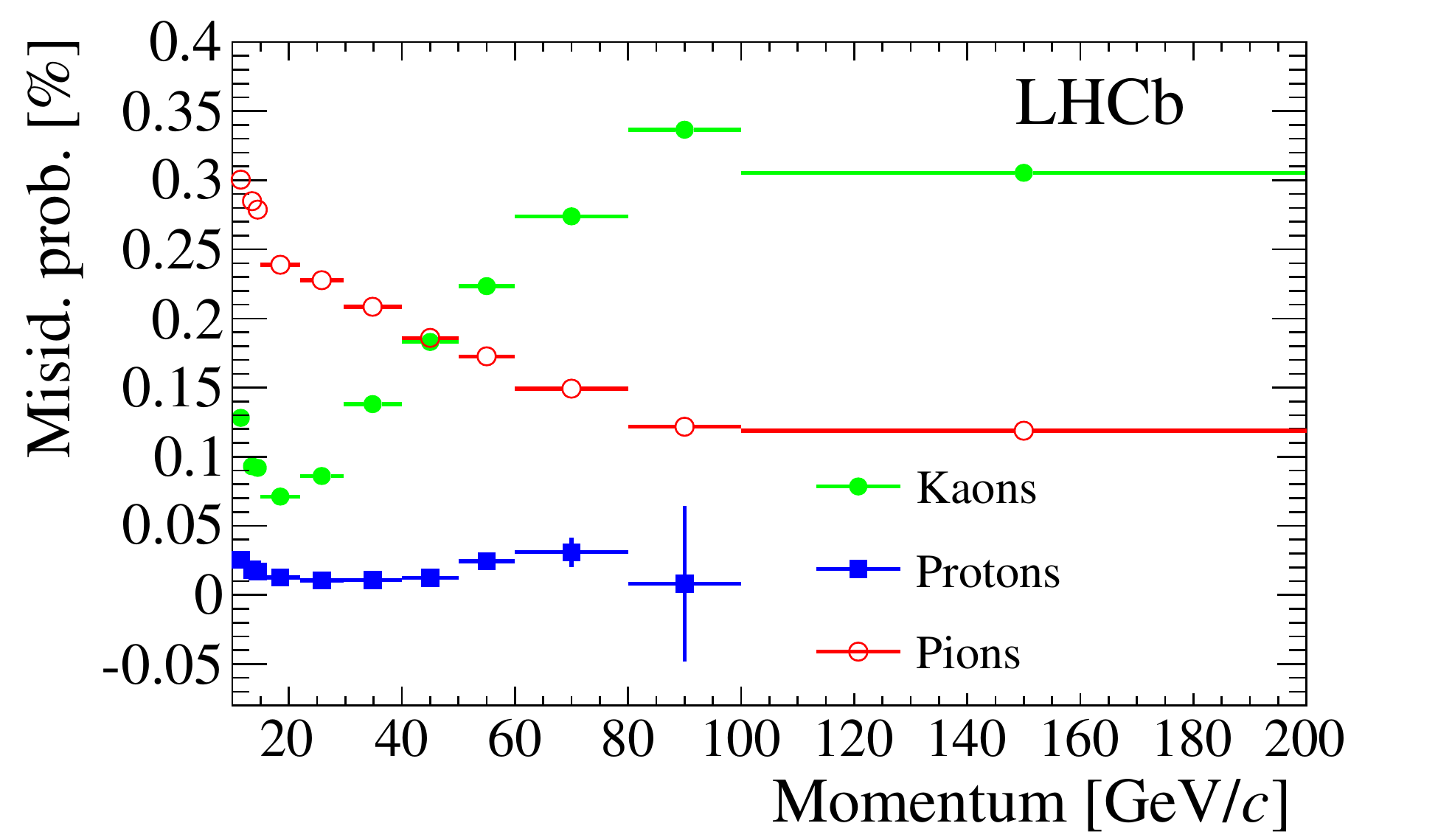}
  \caption{\small \label{fig:misidplots:p}
    Probability for pions, kaons and protons
    to be misidentified as muons, as a function of the particle momentum. 
  }
\end{figure}

The hadron fractions are determined in each bin from fits to the
two-dimensional distribution of the particle identification variables
$\DLL_{K/\pi}$ and $\DLL_{p/\pi}$ in the \jpsi-track sample.
The discriminating power achievable with these variables is illustrated in
Fig.~\ref{fig:planeLikelihoods}.
The misidentification probabilities $W_h$ are obtained by applying the
muon identification criteria to the calibration samples. The result as
a function of momentum, averaged over $\eta$ and $N_t$, is shown in Fig.~\ref{fig:misidplots:p}.
The approximately exponential dependence for pions is due to decays in
flight, while the Cherenkov detectors provide a better identification
performance for low momentum kaons.
The average value of $W$ is found to be 0.20\%,
corresponding to an expected yield of  $10\,978 \pm 110$ candidates, where the
uncertainty is statistical only. The two-dimensional (\pst, \Mjm)
\PDF  is obtained  from the 
\jpsi-track events weighted according to Eq.~\ref{eq:misidW}.
A yield of $1686 \pm 90$ candidates is predicted in the detached region
(defined as $\pst>150$~fs), due to \bquark-hadron decays.
The model is validated by comparing it with 
the prediction from a simulated sample of events containing 
a $B \to \jpsi X$ decay, where $B=\Bp,\Bz,\Bs$.
 The yield and \PDF shape are primarily due to a set of exclusive $B$ meson decays,
the most important ones being $\Bz\to\jpsi\Kstarz$ and  $\Bp\to\jpsi\Kp$.

\begin{figure}[tb]
  \centering
  \includegraphics[width=0.75\textwidth]{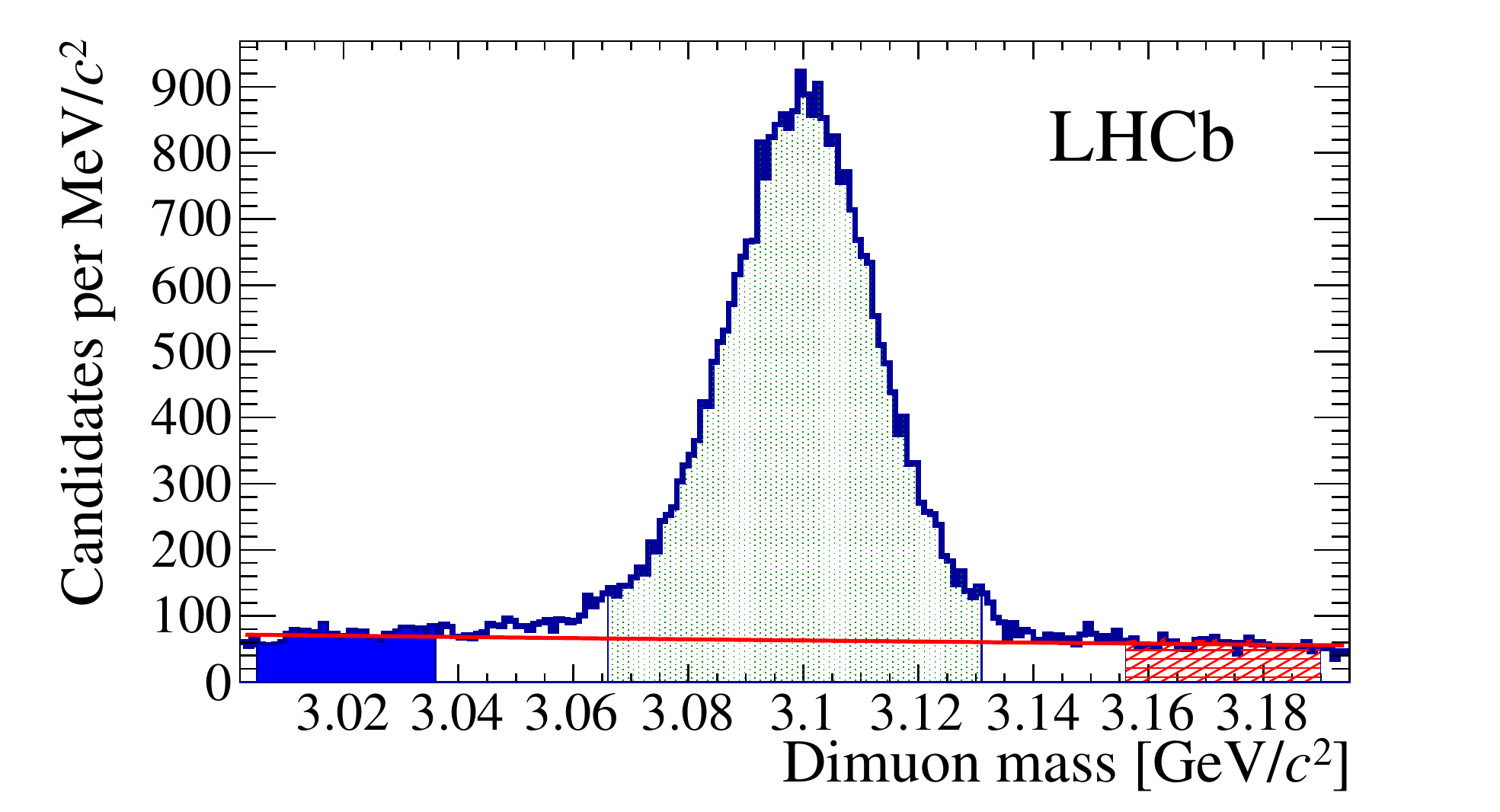}
  \caption{\small Dimuon mass distribution for the \jpsi candidates. The selected
    signal region is shown by the central light-shaded area. The sidebands
    used for the estimation of the fake \jpsi background are shown by the 
    dark-shaded areas, and the function modelling such background by the solid
    red curve.
  }
  \label{fig:fakejpsimass}  
\end{figure}

The fake \jpsi background is modelled using the \jpsi mass sidebands,
as illustrated in Fig.~\ref{fig:fakejpsimass}. The expected yield is
obtained by extrapolating the distribution from the sidebands assuming
an exponential behaviour.
The (\pst, \Mjm) distribution is found to be statistically consistent 
in the two sidebands.
Since the two variables are found to be correlated, a two-dimensional
model is used. To reduce the fluctuations due to  the limited
sample size, a smoothing based on kernel
estimation~\cite{Cranmer:2000du} is
applied to the observed two-dimensional distribution.
The candidates with a fake \jpsi and a misidentified bachelor muon are
already taken into account in the \misbkg category. Their yield
and \PDF shape are estimated with the same technique used  for the \misbkg,
namely by weighting \jpsi-track events in the sideband regions according to
Eq.~\ref{eq:misidW}, and are subtracted from the fake \jpsi model.
After this correction, the fake \jpsi background yield is predicted to be 2994 $\pm$ 109 candidates.

The prompt background component is important for decays close to the PV,
while it is suppressed in the detached region, where most of the signal is expected.
To constrain the effects of the tails of the \pst distribution for prompt background events,
the PV region is included in the fit,  allowing the
yield and shape parameters of the prompt background to be determined from data. 
Alternative fits with a detachment requirement are used as checks for systematic effects. 
The \pst distribution is modelled by a Gaussian 
function, whose parameter values are left free to vary in the fit.
The \Mjm distribution
is obtained from the events in the prompt region, requiring
$-500 < \pst < 10$~fs to remove the signal component, 
making the identification requirements for the bachelor muon more stringent to 
suppress the contamination from the \misbkg. 
Since no correlations are found between \pst and \Mjm in
simulated events, the two-dimensional model is obtained by multiplying the {\PDF}s of
the two variables.

The combinatorial background is modelled using a sample of $18$ million events containing
a  $B \to \jpsi X$ decay, simulated according to the known
\bquark-quark fragmentation  fractions and the $B$ meson branching fractions to
these states, and additional simulated samples of $\Lb\to\jpsi\Lz$ and $\Lb\to\jpsi\proton\Km$
decays to estimate the contribution from \bquark baryons. The measured value of 
the \Lb fragmentation  fraction~\cite{LHCb-PAPER-2011-018} is used, and the inclusive $\Lb\to\jpsi X$
branching fraction is assumed to equal that in 
$B$ meson decays. The modest sample surviving the selection is used to
model the \pst and \Mjm distributions, neglecting their correlation. 
The \pst distribution is parametrised with the sum of two exponential functions,
while the mass distribution is  modelled using the kernel
estimation technique. 
The number of events obtained from simulation is scaled according to the measured
\jpsi production cross-section from \bquark
decays~\cite{LHCb-PAPER-2013-016}, the number of simulated events and
the integrated luminosity of the data sample. The resulting yield is
$974 \pm 168$ candidates, where the uncertainty is statistical. Sizeable systematic
uncertainties are assigned to this simulation-based estimation, as discussed
in Sec.~\ref{sec:System}.

Simulated samples are also used to evaluate possible irreducible
backgrounds from \bquark hadrons (different from \Bc) 
decaying to  $\jpsi\mup X$ final states where all three muons are produced at the 
same vertex.
The only decay mode with a non-negligible contribution is found to
be $\Bs\to\jpsi(\mumu)\phi(\mumu)$, from which fewer
than 20 events are expected.
 This $B\to3\mu$ background represents  only 2\%  of the
 combinatorial background and is merged into that category in the following.

Finally, the background model includes a component to describe events
having an incorrectly associated PV, resulting in a faulty reconstruction of the \pseudot.
These events are modelled by associating the candidates with the
primary vertices measured in the previous selected event. The \PDF
is obtained from two-dimensional kernel estimation smoothing, while
the yield is left free in the fit.

%% file: fit.tex
\section{Fit and results}
\label{sec:Fit}

\begin{figure}[p]
  \centering
\includegraphics[width=.73\textwidth]{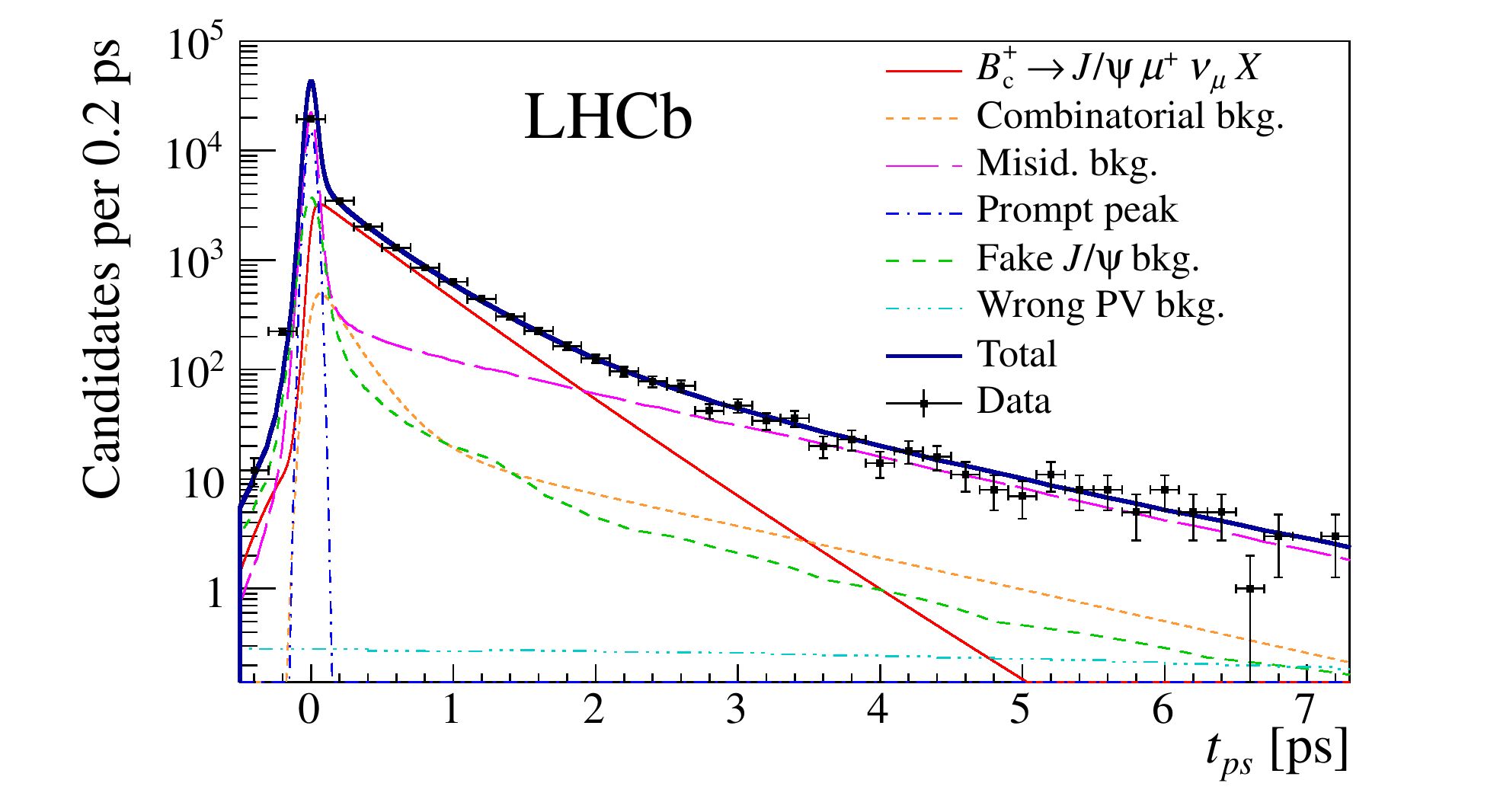}
\put(-200,130){$\mathrm{a})$}\\
\includegraphics[width=.73\textwidth]{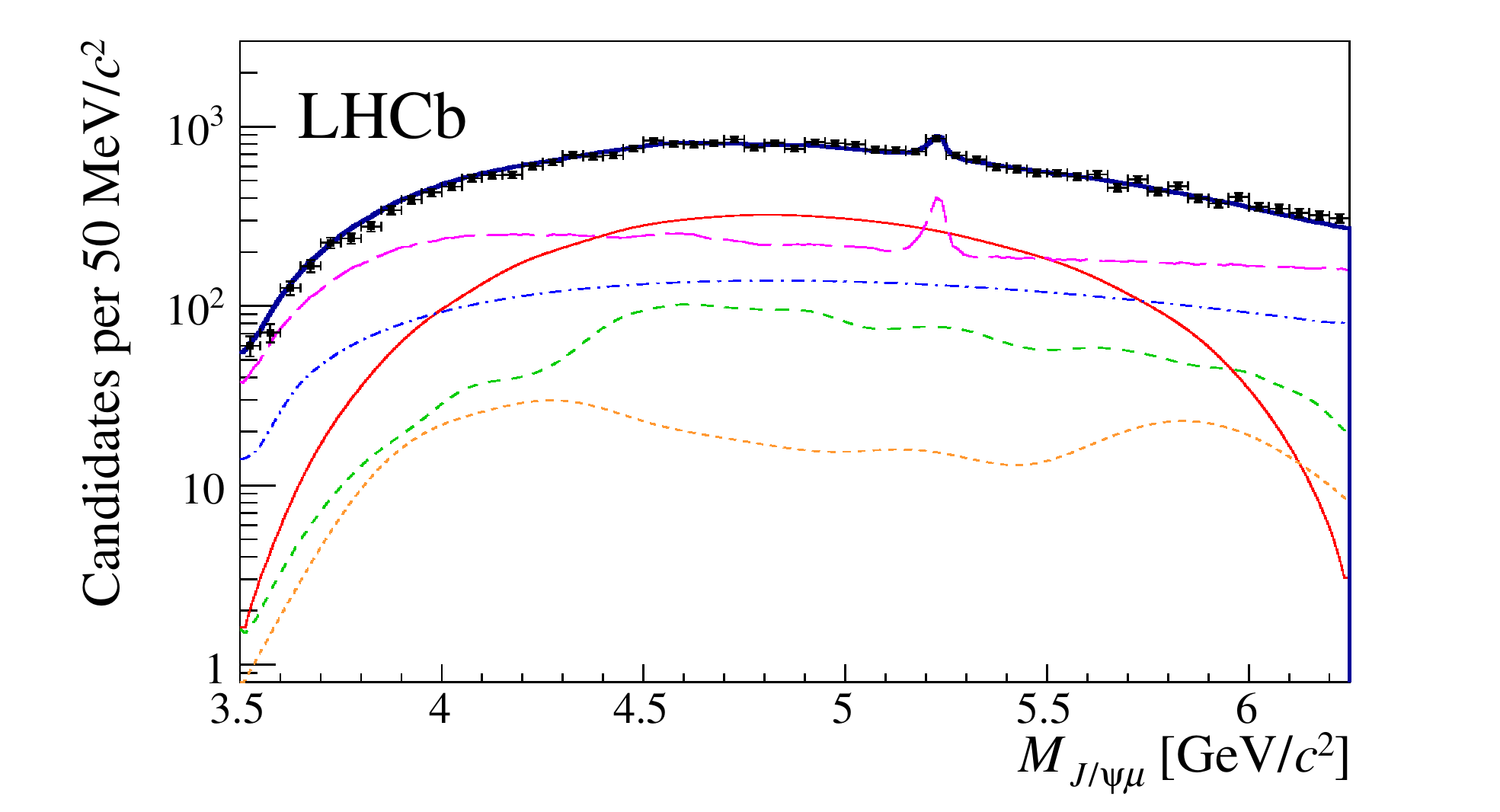}
\put(-60,150){$\mathrm{b})$}\\
\includegraphics[width=.73\textwidth]{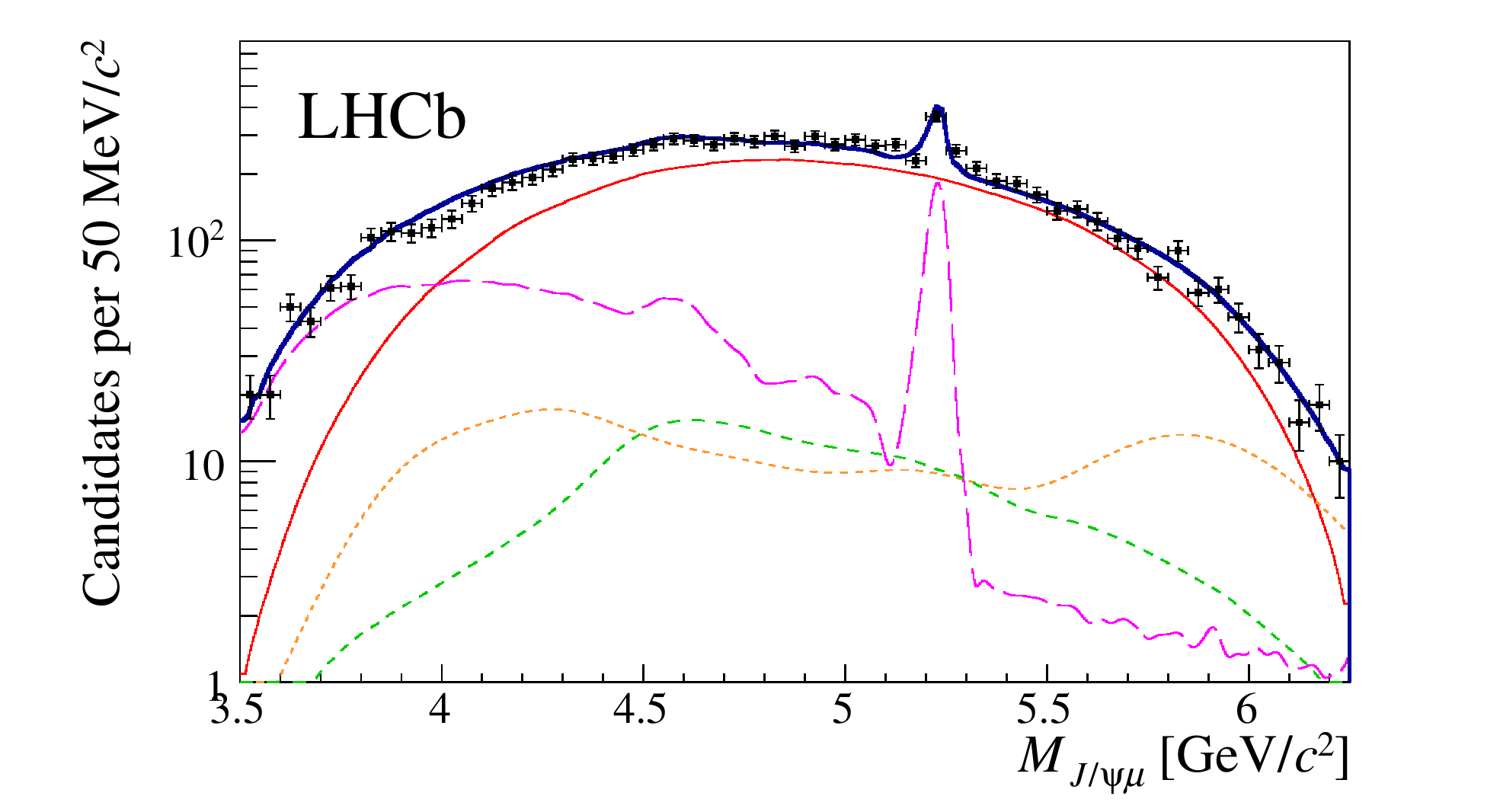}
\put(-60,150){$\mathrm{c})$}\\
  \caption{\small \label{fig:globalfit}
    Result of the two-dimensional fit of the \BctojmX model.
    Projections of the total fit function and its components are shown for
    (a) the \pseudot, (b) the mass of all events, and (c) the mass of the detached events ($\pst > 150$~fs).
  }
\end{figure}

The \Bc lifetime $\tau$ is determined from a maximum likelihood unbinned fit to the (\pst, \Mjm)
distribution of the selected sample, in the range $-1.5 <\pst<8$~ps and
$3.5 < \Mjm < 6.25$~\gevcc.
To avoid inadvertent experimenter bias, an unknown offset
is added to the result for $\tau$, and is removed
only after the finalization of the event selection and analysis procedure.
The other free parameters of the fit are the mean and
width of the \pst resolution function for the prompt background, and
the yields for the signal, the prompt background, and the candidates with an incorrectly
associated PV. The yield parameters for the other background
components are Gaussian-constrained to their predicted values. The
total yield is constrained to the number of events in
the sample. Figure~\ref{fig:globalfit} shows the projected
distributions of the two variables, together with the signal and
background contributions obtained from the fit.  

The fitted number of signal
candidates is $8995 \pm 103$. The \Bc lifetime is determined to be
$\tau= \CENTRALVALUE \pm \STATERROR$~fs, where the uncertainty is statistical only. 
The total number of background candidates is $20\,760 \pm 120$,
of which 2585 have $\pst>150$~fs. 
In the detached region, signal decays dominate the sample,
particularly for \Mjm values above  4.5 \gevcc.
The number of candidates with an incorrectly associated PV is found to be  $12 \pm 5$,
corresponding to a probability of incorrect association smaller than 0.1\%. 
 \afterpage{\clearpage}
The fitted mean and width of the prompt peak are $-2.1 \pm 0.9$~fs
and $32.8 \pm 0.7$~fs, respectively, in excellent agreement with the values
obtained from simulation. The correlations between $\tau$ and the other free parameters 
are all below 20\%. 
Residuals from the fit are consistent with zero in the explored
region of the (\pst, \Mjm) plane.  

A goodness-of-fit test is performed by dividing the region into 100$\times$100
equally sized bins and computing a \chisq from the bins for which
the expected event yield is larger than 0.5. The resulting $p$-value is 0.20.
The method is validated using a set of pseudo-experiments generated
according to the fitted model, where the $p$-value distribution is found to
be consistent with the expected uniform distribution in [0,1].
Tests on pseudo-experiments also show that the fit provides unbiased estimates 
for the lifetime and its statistical uncertainty.

%% file: syst.tex
\section{Systematic uncertainties and checks}
\label{sec:System}

The assigned systematic uncertainties to the \Bc lifetime determination, 
described in the following, are summarised in Table
\ref{tab:systematics}.
Since limited experimental information is available on semileptonic \Bc
decays, uncertainties on the assumed signal \PDF are estimated by constraining
generic model variations using the distributions observed in data, rather
than relying on theoretical predictions.  The \Bc production  spectra
obtained with the \mbox{\textsc{Bcvegpy}} generator are validated using
the measured spectra in $\Bc\to\jpsi\pip$ decays and found to be in
good agreement. Linear deformations are applied to the rapidity
and momentum spectra by reweighting the simulated events.
The fit is repeated after applying the maximum deformations indicated
by the comparison with the  data distributions.
The effect on the lifetime is found to be  within $\pm$1.0~fs.

\begin{figure}[tb]
  \centering
  \includegraphics[width=0.53\textwidth]{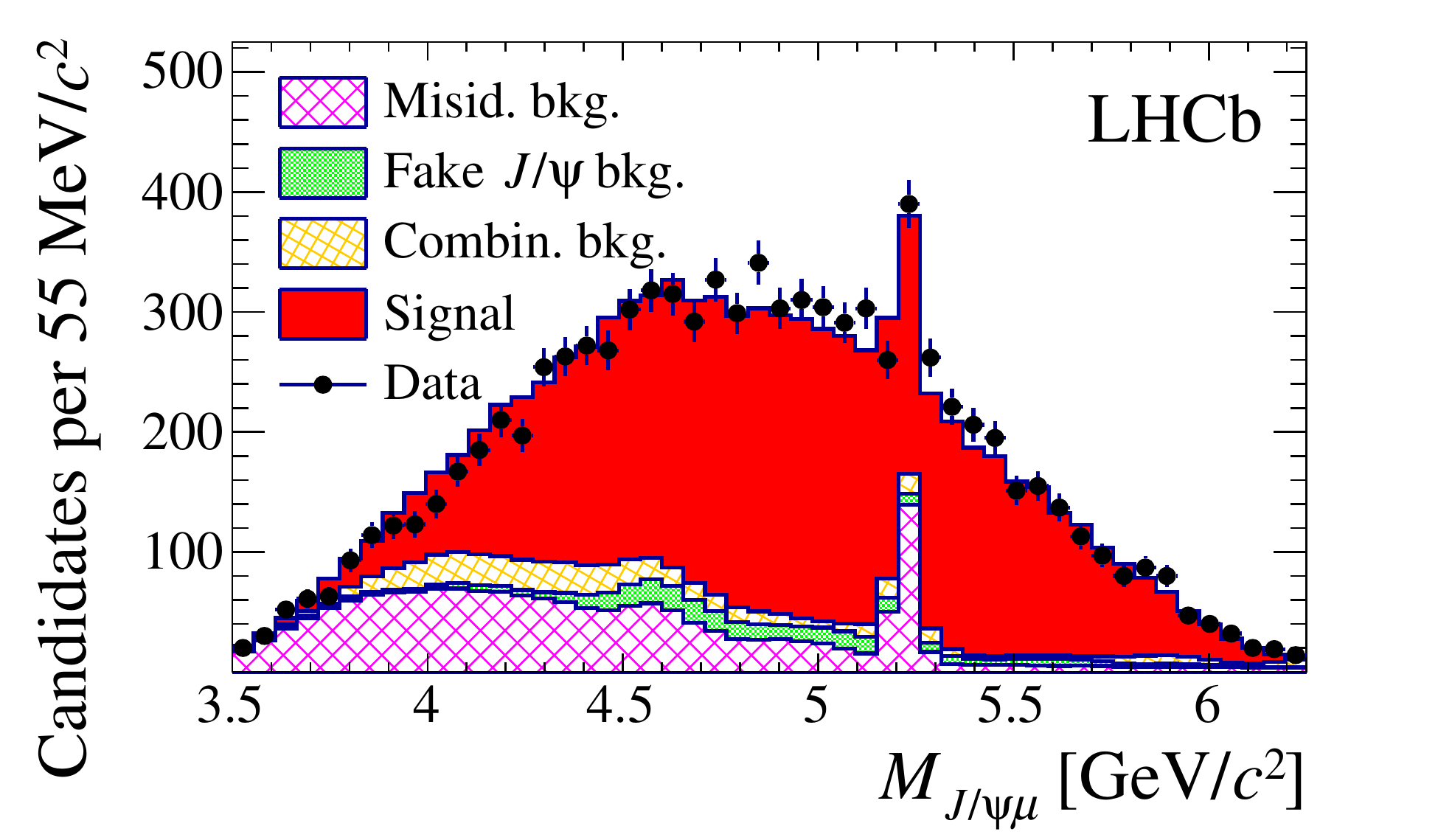}
  \put(-45, 100){$\mathrm{a)}$}\\
  \includegraphics[width=0.49\textwidth]{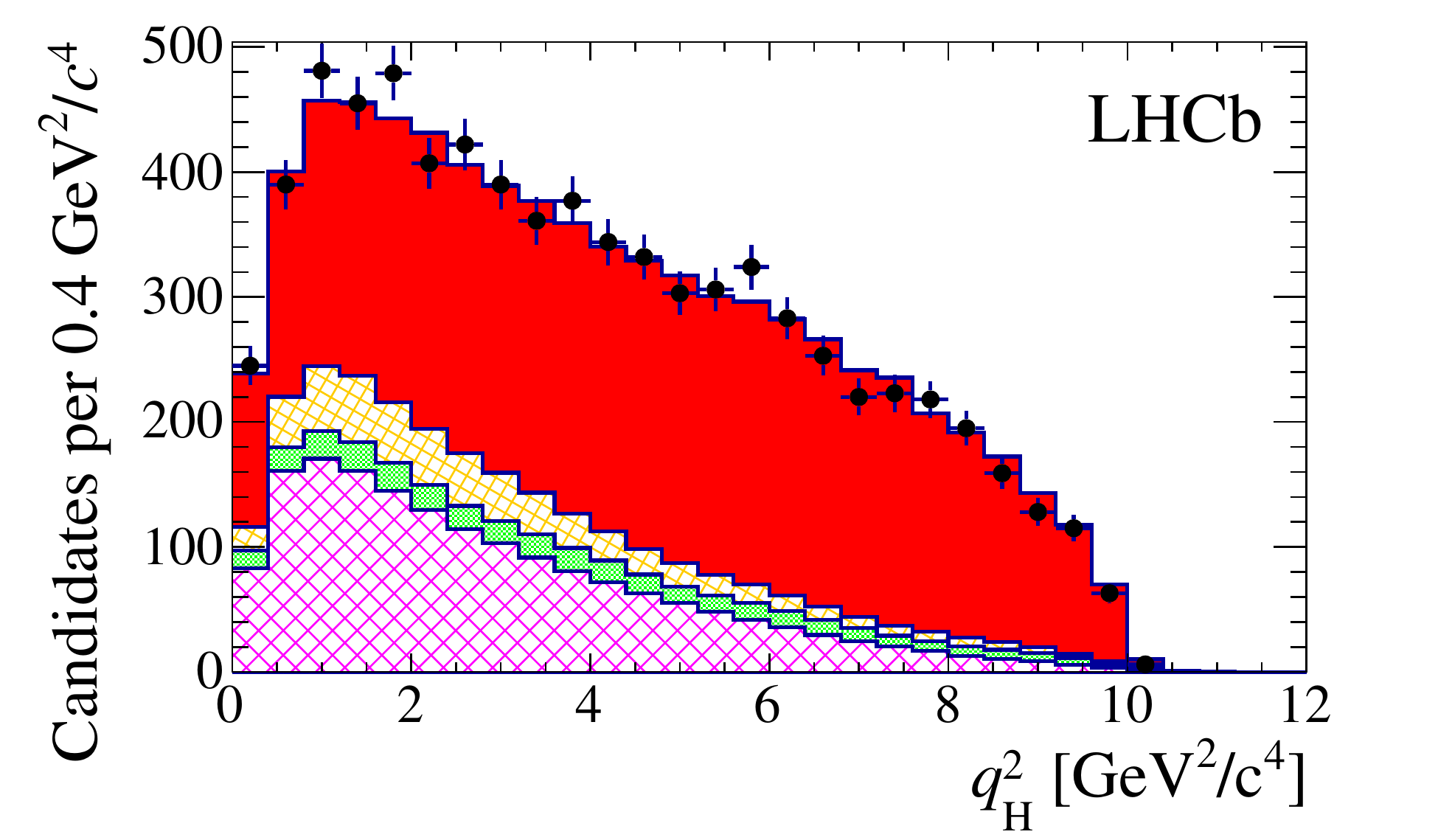}
  \put(-45, 90){$\mathrm{b)}$}
  \includegraphics[width=0.49\textwidth]{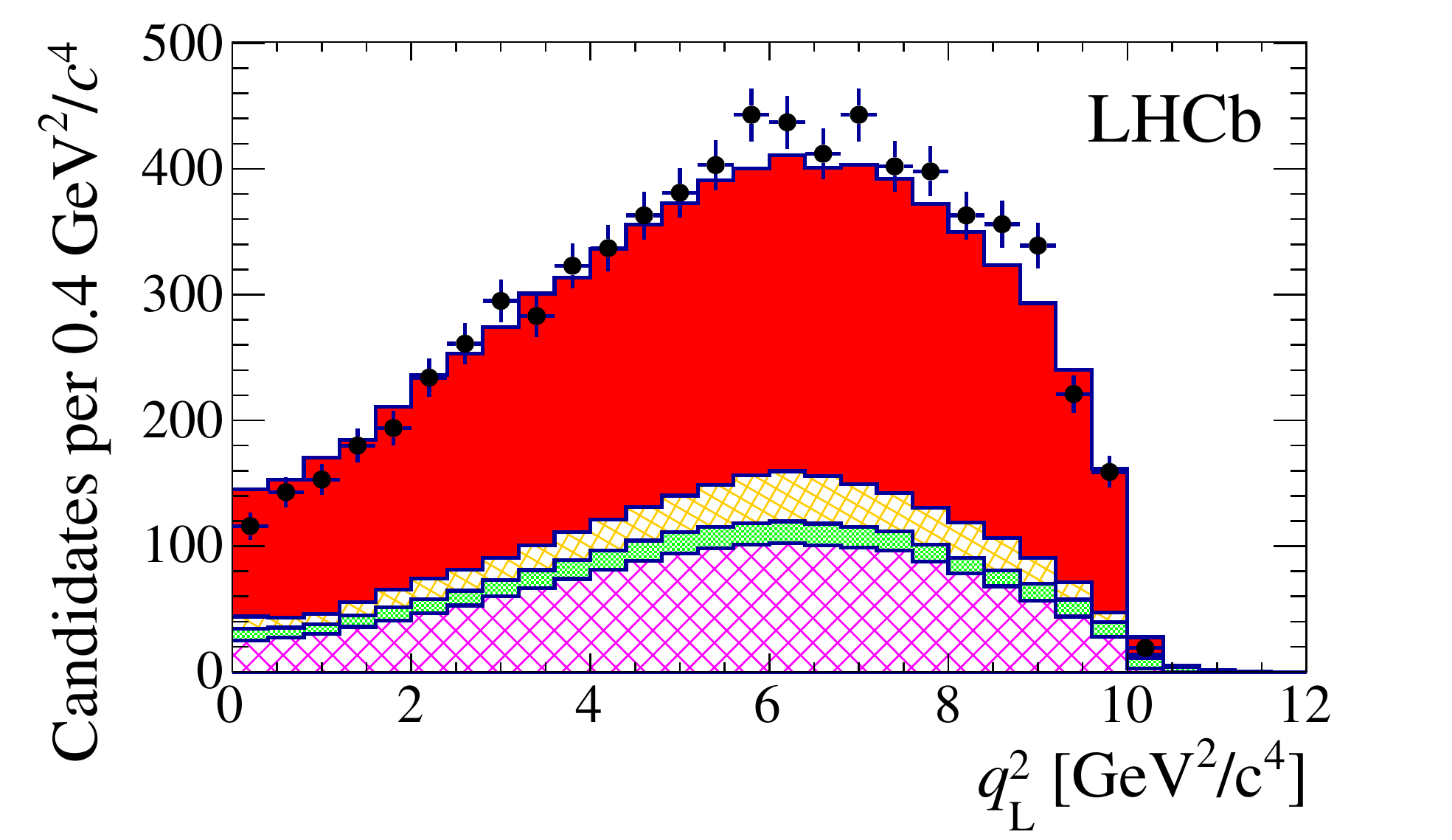}
  \put(-45, 90){$\mathrm{c)}$}\\
  \caption{\small \label{fig:zerodeformation}
    Binned distributions of (a) \Mjm, and (b,c)
    the two $q^2$ solutions for events in the detached region. 
    The modelled contributions for \misbkg (hatched dark violet),
    fake \jpsi background (filled light green), 
    combinatorial background (hatched light orange) and  signal (filled dark red), are shown,
    stacked on each other. Markers representing data are superimposed.
    The  background yields and {\PDF}s are
    obtained with the techniques described  in
    Sec.~\ref{sec:Background}, and only the signal yield is obtained
    from the fit to the data. 
  }
\end{figure}

\begin{figure}[tb]
  \centering
  \includegraphics[width=.7\textwidth]{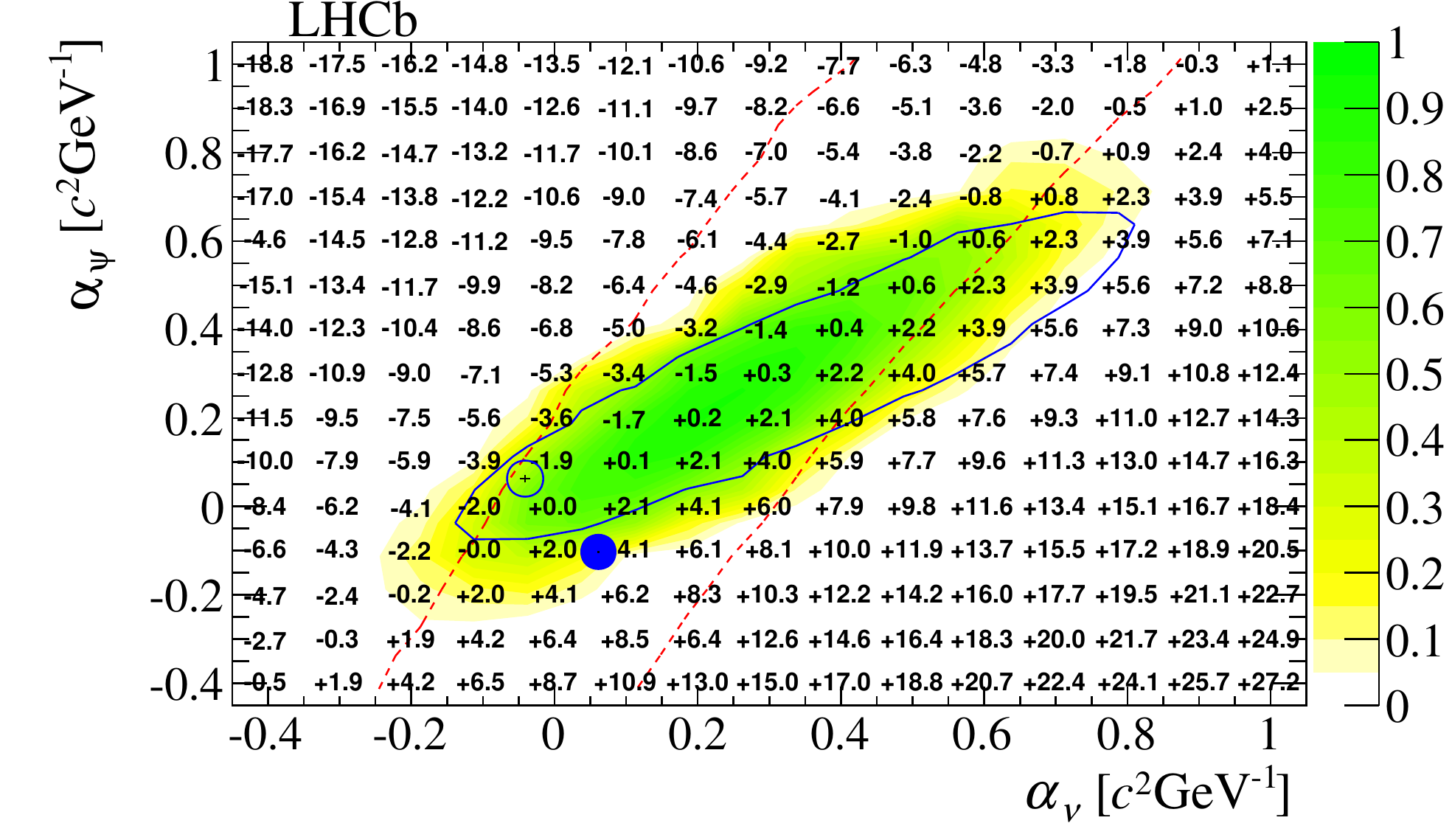}
  \put(-280, 180){$\mathrm{a)}$}\\
  \includegraphics[width=.7\textwidth]{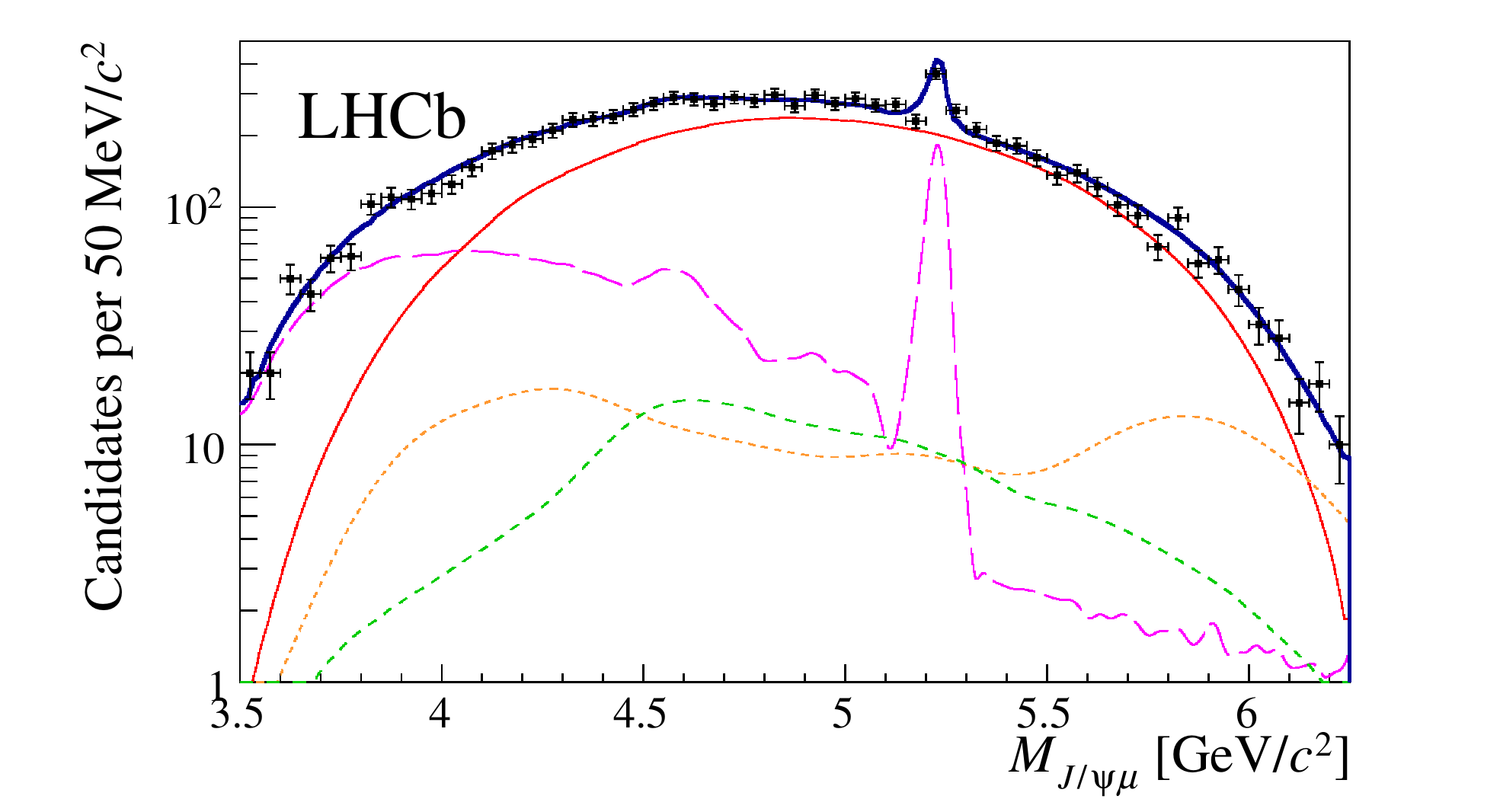}
  \put(-280, 170){$\mathrm{b)}$}\\
  \caption{\small \label{fig:datadrivenmodelsystematic}
      Effect of a generic deformation of the signal model: (a) 
      offset to the lifetime value (expressed in fs) as a function of 
      the deformation parameters; (b) fit projection for the \Mjm
      variable in the detached region after applying
      the deformation maximising the agreement with data 
      ($\alpha_\psi = \alpha_\nu$ = 0.3 $c^2\gev^{-1}$).
      The colour scale on the upper plot indicates the $p$-value of the
      goodness-of-fit test on the three decay kinematic distributions.
      The solid blue (dashed red) lines shows the region having  $p$-value greater than 32\%
      for the \Mjm ($q^2$) test only. The filled (empty) blue marker indicates the deformation parameters  
      that fit best the Ebert (ISGW2) model. The fit components shown on the lower plot follow
      the legend of Fig.~\ref{fig:globalfit}.
    }
\end{figure}

The same technique is used for the uncertainties on the \BctojmX decay
model. A generic model of the distribution in the $\jpsi\mu\nu$ phase space
is defined by applying   the following transformation
 to the nominal model $D(\Mjm^2,\Mmn^2)$
\begin{equation}\label{eq:dalitzdeform}
    D'(\Mjm^2,\Mmn^2) = 
     D(\Mjm^2,\Mmn^2) \times
    \exp(\alpha_\psi \Mjm + \alpha_\nu \Mmn),
\end{equation}
where  $\Mmn^2=q^2$ is the squared mass of the
$\mu\nu$ combination. The deformation parameters $\alpha_\psi$ and $\alpha_\nu$ 
represent generic imperfections of the model for the decay
form factors and feed-down contributions.
The  exponential deformation is chosen to have positive weights 
while keeping an approximately linear deformation  in the
masses for small values of the deformation parameters. 
Partial reconstruction of the decay, using the measured flight direction of the \Bc
meson and its known mass value, is used to determine its momentum  up to  a
two-fold ambiguity.
The agreement between the deformed model and the
data is evaluated, using the signal-enriched detached sample, 
from the distributions of \Mjm and of the 
two $q^2$ solutions $q^2_{\mathrm H}$ ($q^2_{\mathrm L}$) obtained using the higher (lower)
solution for the \Bc momentum. 
The comparison for the nominal model is shown in Fig.~\ref{fig:zerodeformation}.
Figure~\ref{fig:datadrivenmodelsystematic}(a) shows the results of
goodness of fit tests obtained when varying the deformation parameters.
The agreement is assessed by performing a $\chi^2$ test on each of the three 
distributions.
Among the models compatible with data at 90\% confidence level
(combined \mbox{$p$-value $>$ 0.1}), variations of the \Bc lifetime
are within $\pm$5.0~fs, which is assigned as systematic uncertainty. 
It can be noted, by comparing Fig.~\ref{fig:datadrivenmodelsystematic}(b)
with Fig.~\ref{fig:globalfit}(c), that the fit quality in the mass projection
is significantly  improved after applying the deformation that
maximises the combined $p$-value. 

As a consistency check, the model is also varied within the uncertainties
evaluated by comparing available theoretical predictions for the
\Bctojm form factors  and feed-down contributions. 
When the signal model is built using the alternative samples of simulated
\Bctojm decays generated with the Ebert and ISGW2 form-factor models, the   
lifetime changes  by
\offsetUSEEBERT~fs and \offsetUSEISGWTWO~fs, respectively, consistent with the
model-independent evaluation.
Indeed, the deformation parameters corresponding to the best
approximation of the alternative models, shown in
Fig.~\ref{fig:datadrivenmodelsystematic}(a), are  
compatible with data with a confidence level in excess of 90\%.
For the feed-down contributions, the relative decay widths with
respect to the \Bctojm decay are varied according to the range of
values predicted in Refs.~\cite{Kiselev:2002vz,Ebert:2003cn,Wang:2011jt,Ebert:2010zu,Ivanov:2000aj,Colangelo:1999zn,PhysRevD.56.4133,Anisimov:1998uk,PhysRevD.65.014017}.
 More conservatively, each modelled component is
varied in turn by $\pm$100\%.
The maximum  variation with respect to the nominal fit is 0.3~fs.

Several effects concerning the reconstruction of signal events are
considered. 
The resolution model for the signal is varied using a quadruple Gaussian
instead of the nominal triple Gaussian model. The lifetime
variation is \offsetSIGNALMODELFOURGRES~fs, which is assigned as the systematic uncertainty from this 
 source.
The number of \Mjm bins used for the \kf parametrization 
is varied to evaluate the effect of the discretization.
Results obtained with more than ten bins are stable within $\pm 0.1$ fs
and the effect is neglected.

\begin{figure}[tb]
  \centering
  \begin{minipage}{0.49\linewidth}    
    \centering  
    \includegraphics[width=1.1\textwidth]{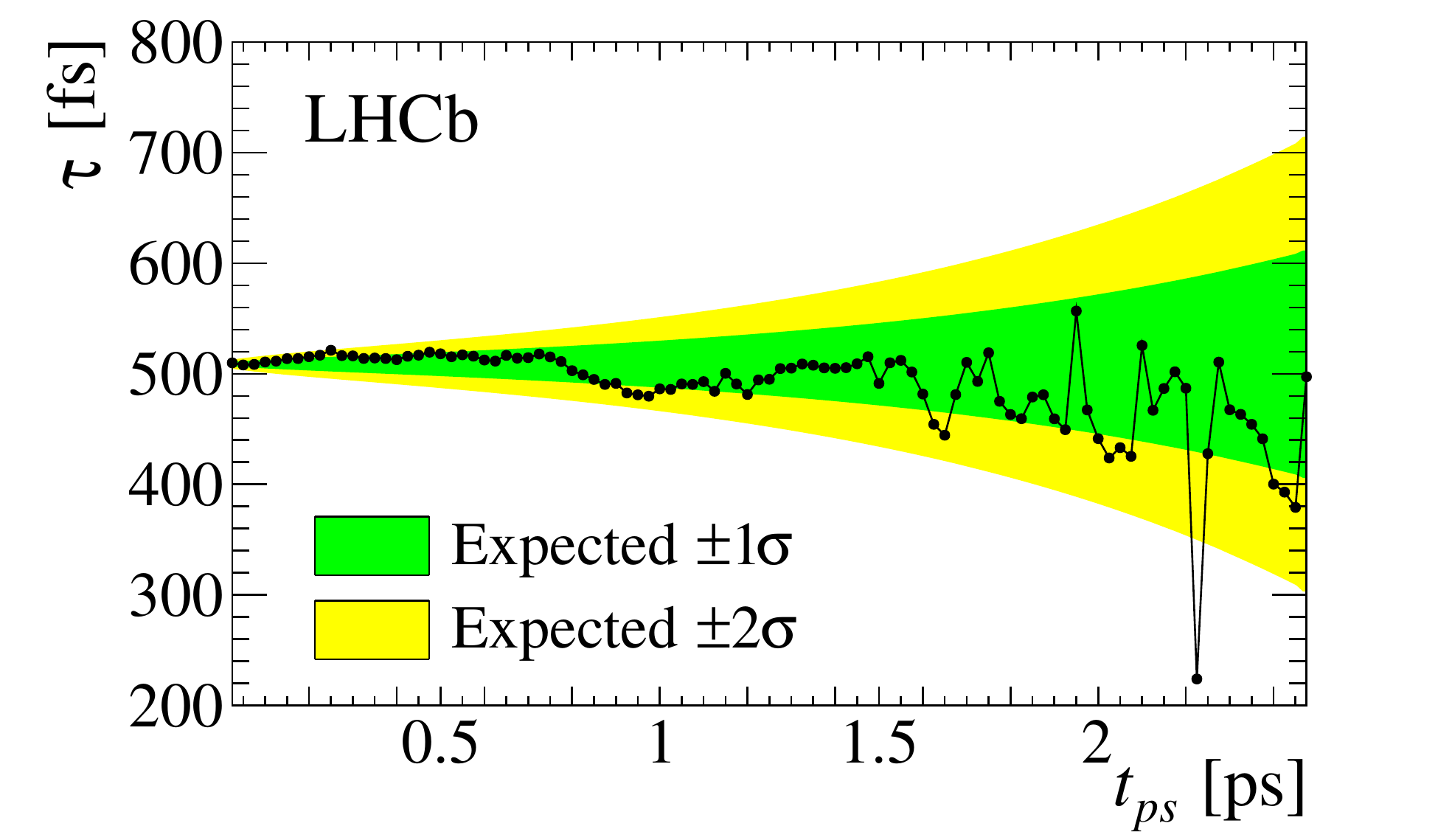}
    \put(-120,119){$\mathrm{a})$}
  \end{minipage}
  \begin{minipage}{0.49\linewidth}
    \centering
    \includegraphics[width=1.1\textwidth]{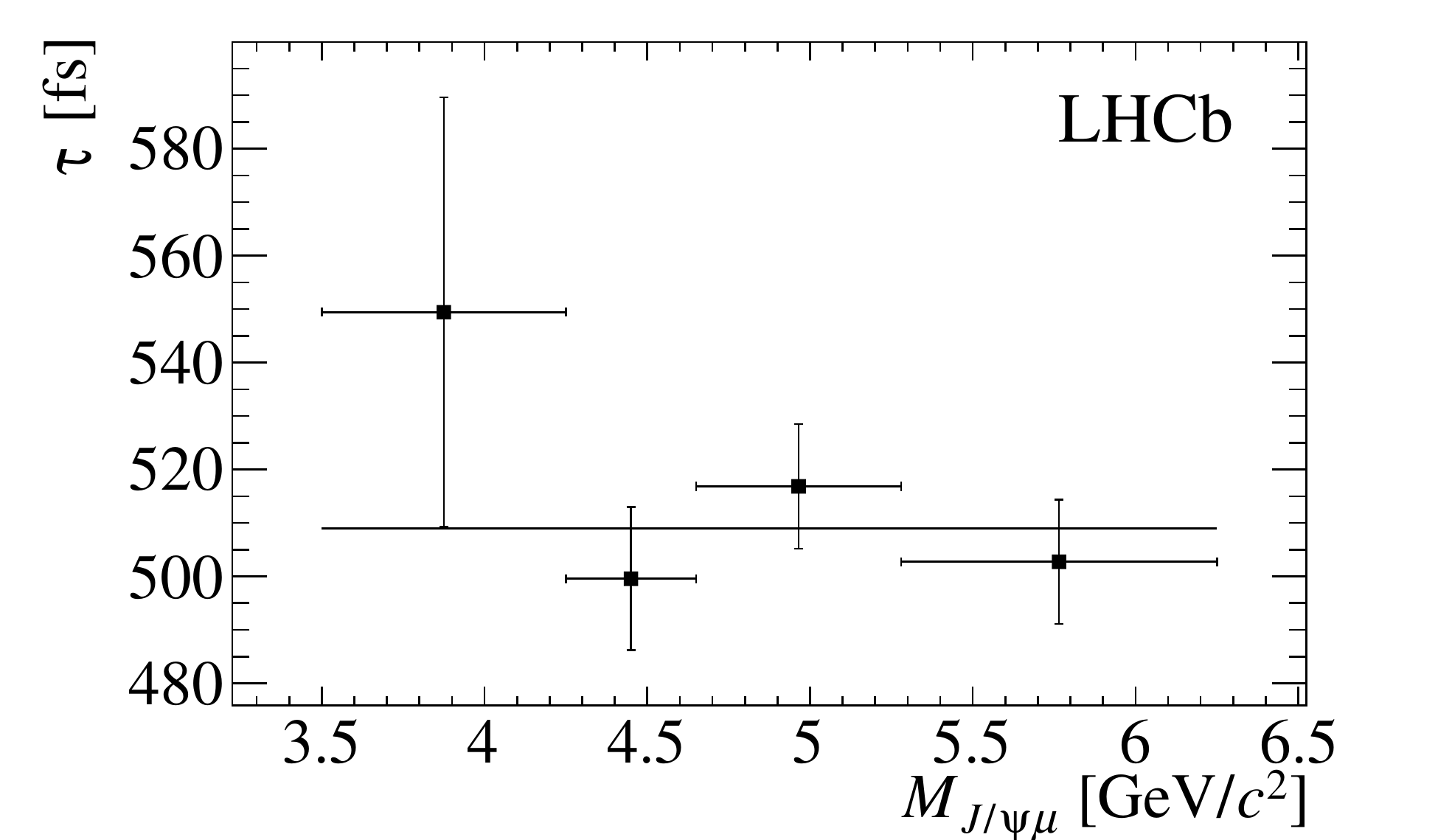}
    \put(-120,119){$\mathrm{b})$}
  \end{minipage}
  \caption{\small  
    Lifetime results obtained after reducing the range of the \pst and \Mjm variables
    (a) removing the events with \pst lower than the threshold reported on the $x$-axis and,
    (b) dividing events in  bins of \Mjm.
    The dark (light) shaded band on (a) shows the expected $\pm$1(2) 
    statistical standard deviation ($\sigma$) of the lifetime variation due to the reduced sample size.  
    The horizontal line on (b) shows the lifetime result of the nominal fit.
  } \label{fig:thresholdpromptcut}\label{fig:massbinfit}
\end{figure}

A possible systematic bias related to the prompt background model is
explored by performing fits with a minimum requirement on the \pst
value. The results, shown in Fig.~\ref{fig:thresholdpromptcut}(a), are
consistent with the expected fluctuations due to the reduced size of the sample.
The lifetime variation obtained with the 
$\pst>150~$fs requirement, which removes most of the prompt candidates, is
\offsetCENTRALVALUENP~fs, corresponding to 1.5 times the expected
statistical error, and  is conservatively taken as the systematic uncertainty.
The fitted lifetime value changes within this uncertainty when  modifying the \pst
resolution function, using a  triple Gaussian shape obtained from simulation instead of
the single  Gaussian with free parameters used in the nominal fit, or
when  using  the  \Mjm distribution predicted using  simulated events with
prompt \jpsi production.

For the fake \jpsi background, the expectation value of its yield is
varied within its systematic uncertainty. The uncertainty on the
\PDF shape is studied using the two alternative models obtained using
only one of the two \jpsi mass sidebands. The observed offsets are
within $\pm\systFAKEJPSINORIGHTSB$~fs.

Since the combinatorial contribution is the only background source whose model relies on simulation,
data-driven checks are performed to evaluate the uncertainty on its predicted yield.
The yield of detached candidates before the bachelor muon identification requirements, which is
expected to be dominated by \bquark-hadron decays, is measured and found to differ by  35\%
from the value predicted by the simulation. To account for a further uncertainty 
related to the efficiency of the muon identification criteria, a systematic uncertainty of  $\pm50\%$ is assigned on
the combinatorial background yield. Another check is performed by comparing the
event yields in data and simulation for candidates with  \Mjm values above the \Bc mass, 
where only combinatorial background is expected.
For this check, an additional requirement $p_T(J/\psi) > 3$~\gevc is
applied on simulated data, since data are filtered   
with such a requirement in this mass region. 
The observed event yield is $221 \pm 14$ events, and the predicted yield is
$201 \pm 73$. The \pseudot and mass distributions are also found to agree.
The quoted uncertainty on the yield corresponds to 
$\pm$\systBBBARNORMINCREASED~fs on $\tau$.
The uncertainty on the \PDF is dominated by the shape of the \pst distribution.
A single exponential rather than  a double exponential function is used, and the
parameters of the nominal function are varied within their statistical
uncertainty. The maximum variation is \offsetBBBARSINGLEEXPONENTIAL~fs.

For the \misbkg, an alternative fit is performed allowing its yield to
vary freely, instead of being Gaussian constrained to its predicted
value. The exercise is repeated using only detached events. The
resulting yields are found to be compatible with the expected ones, 
and the maximum $\tau$ variation of \offsetFREEMISIDNORM~fs is taken
as systematic uncertainty.  The accuracy of the \PDF model is
limited by the size of the calibration and \jpsi-track samples, since
the misidentification probability $W$ of Eq.~\ref{eq:misidW} is parametrised in bins of several variables.
The effect of the uncertainty in each bin is estimated by simulating 1000
alternative {\PDF}s after applying random offsets to the $W$ values, 
according to their statistical uncertainty. The maximum 
variation of the lifetime is  \offsetMISIDDEFORMRB~fs.

Systematic biases on the reconstruction of the \pseudot scale can be
produced by miscalibration of the detector length or momentum scale, and by
a dependence on the decay time of the reconstruction and selection
efficiency $\epsilon$. All these effects have been  evaluated
using simulation and control samples in previous studies.
  The uncertainty on position effects is
known~\cite{LHCb-PAPER-2012-013} to be  dominated by the calibration of
the longitudinal scale. The resulting effect on $\tau$ is within $\pm$1.3~fs.
The momentum scale is varied within its uncertainty~\cite{LHCb-PAPER-2011-035}
and the effect is found to be negligible.
If the dependence of the efficiency $\varepsilon$ on the decay time is
linearly approximated as $\varepsilon(t) \propto (1+\beta t)$, the
bias on the lifetime is about $\beta\tau^2$.
According to the simulation used in this study,  
the value of $\beta$ is compatible with zero within a statistical
uncertainty of 6~ns$^{-1}$. An uncertainty of $\pm 10 $~ns$^{-1}$ on $\beta$
is conservatively assigned, based on data-driven studies of 
the effects contributing to  $\beta$ for some exclusive \bquark-hadron 
to $\jpsi X$ modes~\cite{LHCb-PAPER-2013-065}. The corresponding systematic
uncertainty on $\tau$ is $\pm2.6$~fs.

To estimate the uncertainty on the modelling of events with an incorrectly
associated PV, the fit is repeated removing events where more than one
PV are compatible with the candidate decay. The lifetime changes by  $+1.8$~fs.
A possible effect due to multiple candidates in the same event is
studied by introducing an explicit bias, retaining only the candidate
with the lowest or highest \pst value. The bias is found to be within 1.0~fs.
Finally, the fit procedure is validated using 300 simulated
pseudo-experiments generated according to the nominal fit model.
The average value of the fitted lifetime agrees with the generated
value within the statistical uncertainty of $0.5$~fs.

\begin{table}
  \centering
  \caption{\small \label{tab:systematics}
    Systematic uncertainties on the \Bc lifetime.
  }
  \doSystTable
\end{table}

The sum in quadrature of the mentioned contributions is \SYSTERROR~fs.
Several further consistency checks have been performed to  probe residual
biases not accounted for by the assigned systematic uncertainty.
To check the reliability of the prediction for the \kf
distribution, including the reconstruction effects,
a sample of $\Bz\to\jpsi\Kp\pim$ decays is reconstructed with or without
the pion in the final state.  
Using the information from the fully reconstructed decay, 
the distribution of the \kf, defined in this case as the ratio of
the two reconstructed quantities \pst/$t$, is measured from 
data and compared to the prediction from simulation.
After reweighting for the observed distribution of the  $\jpsi\Kp$ mass,
the distributions are found to agree well, and the average \kf is
predicted to better than 0.1\%, corresponding to
a bias on the lifetime below $0.5$~fs.

In the selected sample, the high-\pst tail of the distribution 
is dominated by \bquark-hadron decays. 
To check for a possible mismodelling of this background, the
analysis is repeated varying the maximum \pst requirement between 2 and 8 ps. 
The resulting lifetime variations are within $\pm$1.5~fs, which is compatible with the
expected statistical fluctuations. 
The fit is also performed in four bins of \Mjm, since the background
and feed-down contributions vary strongly with mass and
become very small above the \Bp mass. As shown in
Fig.~\ref{fig:massbinfit}(b), no significant differences are 
found among the four results. 
Another check performed is to relax the requirement on vertex quality from
 the nominal $\chisq<3$  up to $\chisq<9$. For this test, the yield of 
combinatorial background is allowed to vary. It is found to be
compatible with the expected yield, and to be proportional to the vertex
\chisq threshold, as predicted by the simulation. The corresponding changes in $\tau$ are
within $\pm$2.5~fs and are also compatible with the expected statistical fluctuations. 
Finally, the analysis is repeated after splitting the sample into two parts,
according to the polarity of the spectrometer magnet, which is inverted at regular
intervals during the data taking period.
The difference  between the $\tau$ results from the two  polarities is consistent  with zero  within one standard 
deviation.

%% file: conclusion.tex
\section{Conclusions}\label{sec:Conclusion}
Using \BctojmX semileptonic decays, reconstructed with the \lhcb detector from $pp$ collision data
corresponding to an integrated luminosity of $2\invfb$, the lifetime of the \Bc meson is measured to be
$$ \tau =
\CENTRALVALUEshort\pm\STATERRORshort \stat \pm \SYSTERRORshort \syst~
\text{fs}.$$
This is the most precise measurement of the $B_c^+$ lifetime  to date.
It is consistent with the current world average~\cite{PDG2012} and has less than half 
the uncertainty.
This result will improve the accuracy of most \Bc related
measurements, and provides a means of testing theoretical models
describing the \Bc meson dynamics.
Further improvements are expected from the \lhcb experiment
 using $\Bc\to\jpsi\pip$ decays, where
systematic uncertainties  are expected to be largely uncorrelated with those affecting the present determination.

%% file: acknowledgements.tex
\section*{Acknowledgements}

\noindent We express our gratitude to our colleagues in the CERN
accelerator departments for the excellent performance of the LHC. We
thank the technical and administrative staff at the LHCb
institutes. We acknowledge support from CERN and from the national
agencies: CAPES, CNPq, FAPERJ and FINEP (Brazil); NSFC (China);
CNRS/IN2P3 and Region Auvergne (France); BMBF, DFG, HGF and MPG
(Germany); SFI (Ireland); INFN (Italy); FOM and NWO (The Netherlands);
SCSR (Poland); MEN/IFA (Romania); MinES, Rosatom, RFBR and NRC
``Kurchatov Institute'' (Russia); MinECo, XuntaGal and GENCAT (Spain);
SNSF and SER (Switzerland); NAS Ukraine (Ukraine); STFC (United
Kingdom); NSF (USA). We also acknowledge the support received from the
ERC under FP7. The Tier1 computing centres are supported by IN2P3
(France), KIT and BMBF (Germany), INFN (Italy), NWO and SURF (The
Netherlands), PIC (Spain), GridPP (United Kingdom).
We are indebted to the communities behind the multiple open source software packages we depend on.
We are also thankful for the computing resources and the access to software R\&D tools provided by Yandex LLC (Russia).

%% file: main.bbl
\ifx\mcitethebibliography\mciteundefinedmacro
\PackageError{LHCb.bst}{mciteplus.sty has not been loaded}
{This bibstyle requires the use of the mciteplus package.}\fi
\providecommand{\href}[2]{#2}
\begin{mcitethebibliography}{10}
\mciteSetBstSublistMode{n}
\mciteSetBstMaxWidthForm{subitem}{\alph{mcitesubitemcount})}
\mciteSetBstSublistLabelBeginEnd{\mcitemaxwidthsubitemform\space}
{\relax}{\relax}

\bibitem{Gouz:2002kk}
I.~P. Gouz {\em et~al.}, \ifthenelse{\boolean{articletitles}}{{\it {Prospects
  for the $B_c$ studies at \lhcb}},
  }{}\href{http://dx.doi.org/10.1134/1.1788046}{Phys.\ Atom.\ Nucl.\  {\bf 67}
  (2004) 1559}, \href{http://arxiv.org/abs/hep-ph/0211432}{{\tt
  arXiv:hep-ph/0211432}}\relax
\mciteBstWouldAddEndPuncttrue
\mciteSetBstMidEndSepPunct{\mcitedefaultmidpunct}
{\mcitedefaultendpunct}{\mcitedefaultseppunct}\relax
\EndOfBibitem
\bibitem{LHCb-PAPER-2013-044}
LHCb collaboration, R.~Aaij {\em et~al.},
  \ifthenelse{\boolean{articletitles}}{{\it {Observation of the decay $B_c^+\to
  B_s^0\pi^+$}},
  }{}\href{http://dx.doi.org/10.1103/PhysRevLett.111.181801}{Phys.\ Rev.\
  Lett.\  {\bf 111} (2013) 181801}, \href{http://arxiv.org/abs/1308.4544}{{\tt
  arXiv:1308.4544}}\relax
\mciteBstWouldAddEndPuncttrue
\mciteSetBstMidEndSepPunct{\mcitedefaultmidpunct}
{\mcitedefaultendpunct}{\mcitedefaultseppunct}\relax
\EndOfBibitem
\bibitem{Abe:1998wi}
CDF collaboration, F.~Abe {\em et~al.},
  \ifthenelse{\boolean{articletitles}}{{\it {Observation of the $B_c$ meson in
  $p\bar{p}$ collisions at $\sqrt{s} = 1.8$ TeV}},
  }{}\href{http://dx.doi.org/10.1103/PhysRevLett.81.2432}{Phys.\ Rev.\ Lett.\
  {\bf 81} (1998) 2432}, \href{http://arxiv.org/abs/hep-ex/9805034}{{\tt
  arXiv:hep-ex/9805034}}\relax
\mciteBstWouldAddEndPuncttrue
\mciteSetBstMidEndSepPunct{\mcitedefaultmidpunct}
{\mcitedefaultendpunct}{\mcitedefaultseppunct}\relax
\EndOfBibitem
\bibitem{Beneke:1996xe}
M.~Beneke and G.~Buchalla, \ifthenelse{\boolean{articletitles}}{{\it {$B_c$
  meson lifetime}}, }{}\href{http://dx.doi.org/10.1103/PhysRevD.53.4991}{Phys.\
  Rev.\  {\bf D53} (1996) 4991},
  \href{http://arxiv.org/abs/hep-ph/9601249}{{\tt arXiv:hep-ph/9601249}}\relax
\mciteBstWouldAddEndPuncttrue
\mciteSetBstMidEndSepPunct{\mcitedefaultmidpunct}
{\mcitedefaultendpunct}{\mcitedefaultseppunct}\relax
\EndOfBibitem
\bibitem{Anisimov:1998uk}
A.~Y. Anisimov, I.~M. Narodetsky, C.~Semay, and B.~Silvestre-Brac,
  \ifthenelse{\boolean{articletitles}}{{\it {The $B_c$ meson lifetime in the
  light-front constituent quark model}},
  }{}\href{http://dx.doi.org/10.1016/S0370-2693(99)00273-7}{Phys.\ Lett.\  {\bf
  B452} (1999) 129}, \href{http://arxiv.org/abs/hep-ph/9812514}{{\tt
  arXiv:hep-ph/9812514}}\relax
\mciteBstWouldAddEndPuncttrue
\mciteSetBstMidEndSepPunct{\mcitedefaultmidpunct}
{\mcitedefaultendpunct}{\mcitedefaultseppunct}\relax
\EndOfBibitem
\bibitem{Kiselev:2000pp}
V.~V. Kiselev, A.~E. Kovalsky, and A.~K. Likhoded,
  \ifthenelse{\boolean{articletitles}}{{\it {$B_c$ decays and lifetime in QCD
  sum rules}}, }{}\href{http://dx.doi.org/10.1016/S0550-3213(00)00386-2}{Nucl.\
  Phys.\  {\bf B585} (2000) 353},
  \href{http://arxiv.org/abs/hep-ph/0002127}{{\tt arXiv:hep-ph/0002127}}\relax
\mciteBstWouldAddEndPuncttrue
\mciteSetBstMidEndSepPunct{\mcitedefaultmidpunct}
{\mcitedefaultendpunct}{\mcitedefaultseppunct}\relax
\EndOfBibitem
\bibitem{Chang:2000ac}
C.-H. Chang, S.-L. Chen, T.-F. Feng, and X.-Q. Li,
  \ifthenelse{\boolean{articletitles}}{{\it {Lifetime of the $B_c$ meson and
  some relevant problems}},
  }{}\href{http://dx.doi.org/10.1103/PhysRevD.64.014003}{Phys.\ Rev.\  {\bf
  D64} (2001) 014003}, \href{http://arxiv.org/abs/hep-ph/0007162}{{\tt
  arXiv:hep-ph/0007162}}\relax
\mciteBstWouldAddEndPuncttrue
\mciteSetBstMidEndSepPunct{\mcitedefaultmidpunct}
{\mcitedefaultendpunct}{\mcitedefaultseppunct}\relax
\EndOfBibitem
\bibitem{PDG2012}
Particle Data Group, J.~Beringer {\em et~al.},
  \ifthenelse{\boolean{articletitles}}{{\it {\href{http://pdg.lbl.gov/}{Review
  of particle physics}}},
  }{}\href{http://dx.doi.org/10.1103/PhysRevD.86.010001}{Phys.\ Rev.\  {\bf
  D86} (2012) 010001}, {and 2013 partial update for the 2014 edition}\relax
\mciteBstWouldAddEndPuncttrue
\mciteSetBstMidEndSepPunct{\mcitedefaultmidpunct}
{\mcitedefaultendpunct}{\mcitedefaultseppunct}\relax
\EndOfBibitem
\bibitem{Abulencia:2006zu}
CDF collaboration, A.~Abulencia {\em et~al.},
  \ifthenelse{\boolean{articletitles}}{{\it {Measurement of the \Bc meson
  lifetime using the decay mode $\Bc \to \jpsi\ep\neue$}},
  }{}\href{http://dx.doi.org/10.1103/PhysRevLett.97.012002}{Phys.\ Rev.\ Lett.\
   {\bf 97} (2006) 012002}, \href{http://arxiv.org/abs/hep-ex/0603027}{{\tt
  arXiv:hep-ex/0603027}}\relax
\mciteBstWouldAddEndPuncttrue
\mciteSetBstMidEndSepPunct{\mcitedefaultmidpunct}
{\mcitedefaultendpunct}{\mcitedefaultseppunct}\relax
\EndOfBibitem
\bibitem{Abazov:2008rba}
D0 collaboration, V.~M. Abazov {\em et~al.},
  \ifthenelse{\boolean{articletitles}}{{\it {Measurement of the lifetime of the
  $B_c^\pm$ meson in the semileptonic decay channel}},
  }{}\href{http://dx.doi.org/10.1103/PhysRevLett.102.092001}{Phys.\ Rev.\
  Lett.\  {\bf 102} (2009) 092001}, \href{http://arxiv.org/abs/0805.2614}{{\tt
  arXiv:0805.2614}}\relax
\mciteBstWouldAddEndPuncttrue
\mciteSetBstMidEndSepPunct{\mcitedefaultmidpunct}
{\mcitedefaultendpunct}{\mcitedefaultseppunct}\relax
\EndOfBibitem
\bibitem{Aaltonen:2012yb}
CDF collaboration, T.~Aaltonen {\em et~al.},
  \ifthenelse{\boolean{articletitles}}{{\it {Measurement of the $B_c^{-}$ meson
  lifetime in the decay $B_{c}^{-} \rightarrow J/\psi~\pi^{-}$}},
  }{}\href{http://dx.doi.org/10.1103/PhysRevD.87.011101}{Phys.\ Rev.\  {\bf
  D87} (2013) 011101}, \href{http://arxiv.org/abs/1210.2366}{{\tt
  arXiv:1210.2366}}\relax
\mciteBstWouldAddEndPuncttrue
\mciteSetBstMidEndSepPunct{\mcitedefaultmidpunct}
{\mcitedefaultendpunct}{\mcitedefaultseppunct}\relax
\EndOfBibitem
\bibitem{LHCb-PAPER-2011-044}
LHCb collaboration, R.~Aaij {\em et~al.},
  \ifthenelse{\boolean{articletitles}}{{\it {First observation of the decay
  $\Bc \to \jpsi \pi^+\pi^-\pi^+$}},
  }{}\href{http://dx.doi.org/10.1103/PhysRevLett.108.251802}{Phys.\ Rev.\
  Lett.\  {\bf 108} (2012) 251802}, \href{http://arxiv.org/abs/1204.0079}{{\tt
  arXiv:1204.0079}}\relax
\mciteBstWouldAddEndPuncttrue
\mciteSetBstMidEndSepPunct{\mcitedefaultmidpunct}
{\mcitedefaultendpunct}{\mcitedefaultseppunct}\relax
\EndOfBibitem
\bibitem{LHCb-PAPER-2012-054}
LHCb collaboration, R.~Aaij {\em et~al.},
  \ifthenelse{\boolean{articletitles}}{{\it {Observation of the decay $B^+_c
  \to \psi(2S)\pi^+$}},
  }{}\href{http://dx.doi.org/10.1103/PhysRevD.87.071103}{Phys.\ Rev.\  {\bf
  D87} (2013) 071103(R)}, \href{http://arxiv.org/abs/1303.1737}{{\tt
  arXiv:1303.1737}}\relax
\mciteBstWouldAddEndPuncttrue
\mciteSetBstMidEndSepPunct{\mcitedefaultmidpunct}
{\mcitedefaultendpunct}{\mcitedefaultseppunct}\relax
\EndOfBibitem
\bibitem{LHCb-PAPER-2013-010}
LHCb collaboration, R.~Aaij {\em et~al.},
  \ifthenelse{\boolean{articletitles}}{{\it {Observation of \decay{\Bcp}{\jpsi
  \Dsp} and \decay{\Bcp}{\jpsi \Dssp} decays}},
  }{}\href{http://dx.doi.org/10.1103/PhysRevD.87.112012}{Phys.\ Rev.\  {\bf
  D87} (2013) 112012}, \href{http://arxiv.org/abs/1304.4530}{{\tt
  arXiv:1304.4530}}\relax
\mciteBstWouldAddEndPuncttrue
\mciteSetBstMidEndSepPunct{\mcitedefaultmidpunct}
{\mcitedefaultendpunct}{\mcitedefaultseppunct}\relax
\EndOfBibitem
\bibitem{LHCb-PAPER-2013-021}
LHCb collaboration, R.~Aaij {\em et~al.},
  \ifthenelse{\boolean{articletitles}}{{\it {First observation of the decay
  $B_c^+ \to J/\psi K^+$}},
  }{}\href{http://dx.doi.org/10.1007/JHEP09(2013)075}{JHEP {\bf 09} (2013)
  075}, \href{http://arxiv.org/abs/1306.6723}{{\tt arXiv:1306.6723}}\relax
\mciteBstWouldAddEndPuncttrue
\mciteSetBstMidEndSepPunct{\mcitedefaultmidpunct}
{\mcitedefaultendpunct}{\mcitedefaultseppunct}\relax
\EndOfBibitem
\bibitem{LHCb-PAPER-2013-047}
LHCb collaboration, R.~Aaij {\em et~al.},
  \ifthenelse{\boolean{articletitles}}{{\it {Observation of the decay $\Bc \to
  \jpsi \Kp\Km \pip$}}, }{}\href{http://arxiv.org/abs/1309.0587}{{\tt
  arXiv:1309.0587}}, {to appear in JHEP}\relax
\mciteBstWouldAddEndPuncttrue
\mciteSetBstMidEndSepPunct{\mcitedefaultmidpunct}
{\mcitedefaultendpunct}{\mcitedefaultseppunct}\relax
\EndOfBibitem
\bibitem{LHCb-PAPER-2012-028}
LHCb collaboration, R.~Aaij {\em et~al.},
  \ifthenelse{\boolean{articletitles}}{{\it {Measurements of $B_c^+$ production
  and mass with the $B_c^+ \to J/\psi \pi^+$ decay}},
  }{}\href{http://dx.doi.org/10.1103/PhysRevLett.109.232001}{Phys.\ Rev.\
  Lett.\  {\bf 109} (2012) 232001}, \href{http://arxiv.org/abs/1209.5634}{{\tt
  arXiv:1209.5634}}\relax
\mciteBstWouldAddEndPuncttrue
\mciteSetBstMidEndSepPunct{\mcitedefaultmidpunct}
{\mcitedefaultendpunct}{\mcitedefaultseppunct}\relax
\EndOfBibitem
\bibitem{Alves:2008zz}
LHCb collaboration, A.~A. Alves~Jr. {\em et~al.},
  \ifthenelse{\boolean{articletitles}}{{\it {The \lhcb detector at the LHC}},
  }{}\href{http://dx.doi.org/10.1088/1748-0221/3/08/S08005}{JINST {\bf 3}
  (2008) S08005}\relax
\mciteBstWouldAddEndPuncttrue
\mciteSetBstMidEndSepPunct{\mcitedefaultmidpunct}
{\mcitedefaultendpunct}{\mcitedefaultseppunct}\relax
\EndOfBibitem
\bibitem{LHCb-DP-2012-003}
M.~Adinolfi {\em et~al.}, \ifthenelse{\boolean{articletitles}}{{\it
  {Performance of the \lhcb RICH detector at the LHC}},
  }{}\href{http://dx.doi.org/10.1140/epjc/s10052-013-2431-9}{Eur.\ Phys.\ J.\
  {\bf C73} (2013) 2431}, \href{http://arxiv.org/abs/1211.6759}{{\tt
  arXiv:1211.6759}}\relax
\mciteBstWouldAddEndPuncttrue
\mciteSetBstMidEndSepPunct{\mcitedefaultmidpunct}
{\mcitedefaultendpunct}{\mcitedefaultseppunct}\relax
\EndOfBibitem
\bibitem{LHCb-DP-2012-002}
A.~A. Alves~Jr. {\em et~al.}, \ifthenelse{\boolean{articletitles}}{{\it
  {Performance of the LHCb muon system}},
  }{}\href{http://dx.doi.org/10.1088/1748-0221/8/02/P02022}{JINST {\bf 8}
  (2013) P02022}, \href{http://arxiv.org/abs/1211.1346}{{\tt
  arXiv:1211.1346}}\relax
\mciteBstWouldAddEndPuncttrue
\mciteSetBstMidEndSepPunct{\mcitedefaultmidpunct}
{\mcitedefaultendpunct}{\mcitedefaultseppunct}\relax
\EndOfBibitem
\bibitem{LHCb-DP-2012-004}
R.~Aaij {\em et~al.}, \ifthenelse{\boolean{articletitles}}{{\it {The \lhcb
  trigger and its performance in 2011}},
  }{}\href{http://dx.doi.org/10.1088/1748-0221/8/04/P04022}{JINST {\bf 8}
  (2013) P04022}, \href{http://arxiv.org/abs/1211.3055}{{\tt
  arXiv:1211.3055}}\relax
\mciteBstWouldAddEndPuncttrue
\mciteSetBstMidEndSepPunct{\mcitedefaultmidpunct}
{\mcitedefaultendpunct}{\mcitedefaultseppunct}\relax
\EndOfBibitem
\bibitem{Sjostrand:2006za}
T.~Sj\"{o}strand, S.~Mrenna, and P.~Skands,
  \ifthenelse{\boolean{articletitles}}{{\it {PYTHIA 6.4 physics and manual}},
  }{}\href{http://dx.doi.org/10.1088/1126-6708/2006/05/026}{JHEP {\bf 05}
  (2006) 026}, \href{http://arxiv.org/abs/hep-ph/0603175}{{\tt
  arXiv:hep-ph/0603175}}\relax
\mciteBstWouldAddEndPuncttrue
\mciteSetBstMidEndSepPunct{\mcitedefaultmidpunct}
{\mcitedefaultendpunct}{\mcitedefaultseppunct}\relax
\EndOfBibitem
\bibitem{LHCb-PROC-2010-056}
I.~Belyaev {\em et~al.}, \ifthenelse{\boolean{articletitles}}{{\it {Handling of
  the generation of primary events in \gauss, the \lhcb simulation framework}},
  }{}\href{http://dx.doi.org/10.1109/NSSMIC.2010.5873949}{Nuclear Science
  Symposium Conference Record (NSS/MIC) {\bf IEEE} (2010) 1155}\relax
\mciteBstWouldAddEndPuncttrue
\mciteSetBstMidEndSepPunct{\mcitedefaultmidpunct}
{\mcitedefaultendpunct}{\mcitedefaultseppunct}\relax
\EndOfBibitem
\bibitem{Chang:2003cq}
C.-H. Chang, C.~Driouichi, P.~Eerola, and X.~G. Wu,
  \ifthenelse{\boolean{articletitles}}{{\it {BCVEGPY: an event generator for
  hadronic production of the $B_c$ meson}},
  }{}\href{http://dx.doi.org/10.1016/j.cpc.2004.02.005}{Comput.\ Phys.\
  Commun.\  {\bf 159} (2004) 192},
  \href{http://arxiv.org/abs/hep-ph/0309120}{{\tt arXiv:hep-ph/0309120}}\relax
\mciteBstWouldAddEndPuncttrue
\mciteSetBstMidEndSepPunct{\mcitedefaultmidpunct}
{\mcitedefaultendpunct}{\mcitedefaultseppunct}\relax
\EndOfBibitem
\bibitem{Lange:2001uf}
D.~J. Lange, \ifthenelse{\boolean{articletitles}}{{\it {The EvtGen particle
  decay simulation package}},
  }{}\href{http://dx.doi.org/10.1016/S0168-9002(01)00089-4}{Nucl.\ Instrum.\
  Meth.\  {\bf A462} (2001) 152}\relax
\mciteBstWouldAddEndPuncttrue
\mciteSetBstMidEndSepPunct{\mcitedefaultmidpunct}
{\mcitedefaultendpunct}{\mcitedefaultseppunct}\relax
\EndOfBibitem
\bibitem{Golonka:2005pn}
P.~Golonka and Z.~Was, \ifthenelse{\boolean{articletitles}}{{\it {PHOTOS Monte
  Carlo: a precision tool for QED corrections in $Z$ and $W$ decays}},
  }{}\href{http://dx.doi.org/10.1140/epjc/s2005-02396-4}{Eur.\ Phys.\ J.\  {\bf
  C45} (2006) 97}, \href{http://arxiv.org/abs/hep-ph/0506026}{{\tt
  arXiv:hep-ph/0506026}}\relax
\mciteBstWouldAddEndPuncttrue
\mciteSetBstMidEndSepPunct{\mcitedefaultmidpunct}
{\mcitedefaultendpunct}{\mcitedefaultseppunct}\relax
\EndOfBibitem
\bibitem{Allison:2006ve}
Geant4 collaboration, J.~Allison {\em et~al.},
  \ifthenelse{\boolean{articletitles}}{{\it {Geant4 developments and
  applications}}, }{}\href{http://dx.doi.org/10.1109/TNS.2006.869826}{IEEE
  Trans.\ Nucl.\ Sci.\  {\bf 53} (2006) 270}\relax
\mciteBstWouldAddEndPuncttrue
\mciteSetBstMidEndSepPunct{\mcitedefaultmidpunct}
{\mcitedefaultendpunct}{\mcitedefaultseppunct}\relax
\EndOfBibitem
\bibitem{Agostinelli:2002hh}
Geant4 collaboration, S.~Agostinelli {\em et~al.},
  \ifthenelse{\boolean{articletitles}}{{\it {Geant4: a simulation toolkit}},
  }{}\href{http://dx.doi.org/10.1016/S0168-9002(03)01368-8}{Nucl.\ Instrum.\
  Meth.\  {\bf A506} (2003) 250}\relax
\mciteBstWouldAddEndPuncttrue
\mciteSetBstMidEndSepPunct{\mcitedefaultmidpunct}
{\mcitedefaultendpunct}{\mcitedefaultseppunct}\relax
\EndOfBibitem
\bibitem{LHCb-PROC-2011-006}
M.~Clemencic {\em et~al.}, \ifthenelse{\boolean{articletitles}}{{\it {The \lhcb
  simulation application, \gauss: design, evolution and experience}},
  }{}\href{http://dx.doi.org/10.1088/1742-6596/331/3/032023}{{J.\ Phys.\ Conf.\
  Ser.\ } {\bf 331} (2011) 032023}\relax
\mciteBstWouldAddEndPuncttrue
\mciteSetBstMidEndSepPunct{\mcitedefaultmidpunct}
{\mcitedefaultendpunct}{\mcitedefaultseppunct}\relax
\EndOfBibitem
\bibitem{kalman}
R.~E. Kalman, \ifthenelse{\boolean{articletitles}}{{\it A new approach to
  linear filtering and prediction problems}, }{}J.\ Basic Eng.\  {\bf 82}
  (1960) 35\relax
\mciteBstWouldAddEndPuncttrue
\mciteSetBstMidEndSepPunct{\mcitedefaultmidpunct}
{\mcitedefaultendpunct}{\mcitedefaultseppunct}\relax
\EndOfBibitem
\bibitem{Kiselev:1993ea}
V.~V. Kiselev and A.~V. Tkabladze, \ifthenelse{\boolean{articletitles}}{{\it
  {Semileptonic $B_c$ decays from QCD sum rules}},
  }{}\href{http://dx.doi.org/10.1103/PhysRevD.48.5208}{Phys.\ Rev.\  {\bf D48}
  (1993) 5208}\relax
\mciteBstWouldAddEndPuncttrue
\mciteSetBstMidEndSepPunct{\mcitedefaultmidpunct}
{\mcitedefaultendpunct}{\mcitedefaultseppunct}\relax
\EndOfBibitem
\bibitem{Kiselev:1999sc}
V.~V. Kiselev, A.~K. Likhoded, and A.~I. Onishchenko,
  \ifthenelse{\boolean{articletitles}}{{\it {Semileptonic $B_c$ meson decays in
  sum rules of QCD and NRQCD}},
  }{}\href{http://dx.doi.org/10.1016/S0550-3213(99)00505-2}{Nucl.\ Phys.\  {\bf
  B569} (2000) 473}, \href{http://arxiv.org/abs/hep-ph/9905359}{{\tt
  arXiv:hep-ph/9905359}}\relax
\mciteBstWouldAddEndPuncttrue
\mciteSetBstMidEndSepPunct{\mcitedefaultmidpunct}
{\mcitedefaultendpunct}{\mcitedefaultseppunct}\relax
\EndOfBibitem
\bibitem{Kiselev:2002vz}
V.~V. Kiselev, \ifthenelse{\boolean{articletitles}}{{\it {Exclusive decays and
  lifetime of $B_c$ meson in QCD sum rules}},
  }{}\href{http://arxiv.org/abs/hep-ph/0211021}{{\tt
  arXiv:hep-ph/0211021}}\relax
\mciteBstWouldAddEndPuncttrue
\mciteSetBstMidEndSepPunct{\mcitedefaultmidpunct}
{\mcitedefaultendpunct}{\mcitedefaultseppunct}\relax
\EndOfBibitem
\bibitem{Ebert:2003cn}
D.~Ebert, R.~N. Faustov, and V.~O. Galkin,
  \ifthenelse{\boolean{articletitles}}{{\it {Weak decays of the $B_c$ meson to
  charmonium and $D$ mesons in the relativistic quark model}},
  }{}\href{http://dx.doi.org/10.1103/PhysRevD.68.094020}{Phys.\ Rev.\  {\bf
  D68} (2003) 094020}, \href{http://arxiv.org/abs/hep-ph/0306306}{{\tt
  arXiv:hep-ph/0306306}}\relax
\mciteBstWouldAddEndPuncttrue
\mciteSetBstMidEndSepPunct{\mcitedefaultmidpunct}
{\mcitedefaultendpunct}{\mcitedefaultseppunct}\relax
\EndOfBibitem
\bibitem{PhysRevD.52.2783}
D.~Scora and N.~Isgur, \ifthenelse{\boolean{articletitles}}{{\it Semileptonic
  meson decays in the quark model: an update},
  }{}\href{http://dx.doi.org/10.1103/PhysRevD.52.2783}{Phys.\ Rev.\  {\bf D52}
  (1995) 2783}\relax
\mciteBstWouldAddEndPuncttrue
\mciteSetBstMidEndSepPunct{\mcitedefaultmidpunct}
{\mcitedefaultendpunct}{\mcitedefaultseppunct}\relax
\EndOfBibitem
\bibitem{Wang:2011jt}
Z.-h. Wang, G.-L. Wang, and C.-H. Chang,
  \ifthenelse{\boolean{articletitles}}{{\it {The $B_c$ decays to a $P$-wave
  charmonium by the improved Bethe-Salpeter approach}},
  }{}\href{http://dx.doi.org/10.1088/0954-3899/39/1/015009}{J.\ Phys.\  {\bf
  G39} (2012) 015009}, \href{http://arxiv.org/abs/1107.0474}{{\tt
  arXiv:1107.0474}}\relax
\mciteBstWouldAddEndPuncttrue
\mciteSetBstMidEndSepPunct{\mcitedefaultmidpunct}
{\mcitedefaultendpunct}{\mcitedefaultseppunct}\relax
\EndOfBibitem
\bibitem{Ebert:2010zu}
D.~Ebert, R.~N. Faustov, and V.~O. Galkin,
  \ifthenelse{\boolean{articletitles}}{{\it {Semileptonic and nonleptonic
  decays of $B_c$ mesons to orbitally excited heavy mesons in the relativistic
  quark model}}, }{}\href{http://dx.doi.org/10.1103/PhysRevD.82.034019}{Phys.\
  Rev.\  {\bf D82} (2010) 034019}, \href{http://arxiv.org/abs/1007.1369}{{\tt
  arXiv:1007.1369}}\relax
\mciteBstWouldAddEndPuncttrue
\mciteSetBstMidEndSepPunct{\mcitedefaultmidpunct}
{\mcitedefaultendpunct}{\mcitedefaultseppunct}\relax
\EndOfBibitem
\bibitem{Kuang:1988bz}
Y.-P. Kuang, S.~F. Tuan, and T.-M. Yan,
  \ifthenelse{\boolean{articletitles}}{{\it {Hadronic transitions and $^1P_1$
  states of heavy quarkonia}},
  }{}\href{http://dx.doi.org/10.1103/PhysRevD.37.1210}{Phys.\ Rev.\  {\bf D37}
  (1988) 1210}\relax
\mciteBstWouldAddEndPuncttrue
\mciteSetBstMidEndSepPunct{\mcitedefaultmidpunct}
{\mcitedefaultendpunct}{\mcitedefaultseppunct}\relax
\EndOfBibitem
\bibitem{Andreotti:2005vu}
Fermilab E835 collaboration, M.~Andreotti {\em et~al.},
  \ifthenelse{\boolean{articletitles}}{{\it Results of a search for the
  ${h}_{c}(^{1}{P}_{1})$ state of charmonium in the $\eta_{c}\gamma$ and
  $\jpsi\piz$ decay modes},
  }{}\href{http://dx.doi.org/10.1103/PhysRevD.72.032001}{Phys.\ Rev.\  {\bf
  D72} (2005) 032001}\relax
\mciteBstWouldAddEndPuncttrue
\mciteSetBstMidEndSepPunct{\mcitedefaultmidpunct}
{\mcitedefaultendpunct}{\mcitedefaultseppunct}\relax
\EndOfBibitem
\bibitem{Cranmer:2000du}
K.~S. Cranmer, \ifthenelse{\boolean{articletitles}}{{\it {Kernel estimation in
  high-energy physics}},
  }{}\href{http://dx.doi.org/10.1016/S0010-4655(00)00243-5}{Comput.\ Phys.\
  Commun.\  {\bf 136} (2001) 198},
  \href{http://arxiv.org/abs/hep-ex/0011057}{{\tt arXiv:hep-ex/0011057}}\relax
\mciteBstWouldAddEndPuncttrue
\mciteSetBstMidEndSepPunct{\mcitedefaultmidpunct}
{\mcitedefaultendpunct}{\mcitedefaultseppunct}\relax
\EndOfBibitem
\bibitem{LHCb-PAPER-2011-018}
LHCb collaboration, R.~Aaij {\em et~al.},
  \ifthenelse{\boolean{articletitles}}{{\it {Measurement of \bquark hadron
  production fractions in $7\tev$ $pp$ collisions}},
  }{}\href{http://dx.doi.org/10.1103/PhysRevD.85.032008}{Phys.\ Rev.\  {\bf
  D85} (2012) 032008}, \href{http://arxiv.org/abs/1111.2357}{{\tt
  arXiv:1111.2357}}\relax
\mciteBstWouldAddEndPuncttrue
\mciteSetBstMidEndSepPunct{\mcitedefaultmidpunct}
{\mcitedefaultendpunct}{\mcitedefaultseppunct}\relax
\EndOfBibitem
\bibitem{LHCb-PAPER-2013-016}
LHCb collaboration, R.~Aaij {\em et~al.},
  \ifthenelse{\boolean{articletitles}}{{\it {Production of $J/\psi$ and
  $\Upsilon$ mesons in $pp$ collisions at $\sqrt{s}=8 \tev$}},
  }{}\href{http://dx.doi.org/10.1007/JHEP06(2013)064}{JHEP {\bf 06} (2013) 64},
  \href{http://arxiv.org/abs/1304.6977}{{\tt arXiv:1304.6977}}\relax
\mciteBstWouldAddEndPuncttrue
\mciteSetBstMidEndSepPunct{\mcitedefaultmidpunct}
{\mcitedefaultendpunct}{\mcitedefaultseppunct}\relax
\EndOfBibitem
\bibitem{Ivanov:2000aj}
M.~A. Ivanov, J.~G. Korner, and P.~Santorelli,
  \ifthenelse{\boolean{articletitles}}{{\it {Semileptonic decays of the $B_c$
  meson}}, }{}\href{http://dx.doi.org/10.1103/PhysRevD.63.074010}{Phys.\ Rev.\
  {\bf D63} (2001) 074010}, \href{http://arxiv.org/abs/hep-ph/0007169}{{\tt
  arXiv:hep-ph/0007169}}\relax
\mciteBstWouldAddEndPuncttrue
\mciteSetBstMidEndSepPunct{\mcitedefaultmidpunct}
{\mcitedefaultendpunct}{\mcitedefaultseppunct}\relax
\EndOfBibitem
\bibitem{Colangelo:1999zn}
P.~Colangelo and F.~De~Fazio, \ifthenelse{\boolean{articletitles}}{{\it {Using
  heavy quark spin symmetry in semileptonic $B_c$ decays}},
  }{}\href{http://dx.doi.org/10.1103/PhysRevD.61.034012}{Phys.\ Rev.\  {\bf
  D61} (2000) 034012}, \href{http://arxiv.org/abs/hep-ph/9909423}{{\tt
  arXiv:hep-ph/9909423}}\relax
\mciteBstWouldAddEndPuncttrue
\mciteSetBstMidEndSepPunct{\mcitedefaultmidpunct}
{\mcitedefaultendpunct}{\mcitedefaultseppunct}\relax
\EndOfBibitem
\bibitem{PhysRevD.56.4133}
J.-F. Liu and K.-T. Chao, \ifthenelse{\boolean{articletitles}}{{\it ${B}_{c}$
  meson weak decays and $\mathrm{CP}$ violation},
  }{}\href{http://dx.doi.org/10.1103/PhysRevD.56.4133}{Phys.\ Rev.\  {\bf D56}
  (1997) 4133}\relax
\mciteBstWouldAddEndPuncttrue
\mciteSetBstMidEndSepPunct{\mcitedefaultmidpunct}
{\mcitedefaultendpunct}{\mcitedefaultseppunct}\relax
\EndOfBibitem
\bibitem{PhysRevD.65.014017}
C.-H. Chang, Y.-Q. Chen, G.-L. Wang, and H.-S. Zong,
  \ifthenelse{\boolean{articletitles}}{{\it Decays of the meson ${B}_{c}$ to a
  \textit{P}-wave charmonium state $\chi_{c}$ or ${h}_{c}$},
  }{}\href{http://dx.doi.org/10.1103/PhysRevD.65.014017}{Phys.\ Rev.\  {\bf
  D65} (2001) 014017}, \href{http://arxiv.org/abs/hep-ph/0103036}{{\tt
  arXiv:hep-ph/0103036}}\relax
\mciteBstWouldAddEndPuncttrue
\mciteSetBstMidEndSepPunct{\mcitedefaultmidpunct}
{\mcitedefaultendpunct}{\mcitedefaultseppunct}\relax
\EndOfBibitem
\bibitem{LHCb-PAPER-2012-013}
LHCb collaboration, R.~Aaij {\em et~al.},
  \ifthenelse{\boolean{articletitles}}{{\it {Measurement of the effective
  $B_s^0 \to K^+ K^-$ lifetime}},
  }{}\href{http://dx.doi.org/10.1016/j.physletb.2012.08.033}{Phys.\ Lett.\
  {\bf B716} (2012) 393}, \href{http://arxiv.org/abs/1207.5993}{{\tt
  arXiv:1207.5993}}\relax
\mciteBstWouldAddEndPuncttrue
\mciteSetBstMidEndSepPunct{\mcitedefaultmidpunct}
{\mcitedefaultendpunct}{\mcitedefaultseppunct}\relax
\EndOfBibitem
\bibitem{LHCb-PAPER-2011-035}
LHCb collaboration, R.~Aaij {\em et~al.},
  \ifthenelse{\boolean{articletitles}}{{\it {Measurement of \bquark-hadron
  masses}}, }{}\href{http://dx.doi.org/10.1016/j.physletb.2012.01.058}{Phys.\
  Lett.\  {\bf B708} (2012) 241}, \href{http://arxiv.org/abs/1112.4896}{{\tt
  arXiv:1112.4896}}\relax
\mciteBstWouldAddEndPuncttrue
\mciteSetBstMidEndSepPunct{\mcitedefaultmidpunct}
{\mcitedefaultendpunct}{\mcitedefaultseppunct}\relax
\EndOfBibitem
\bibitem{LHCb-PAPER-2013-065}
LHCb collaboration, R.~Aaij {\em et~al.},
  \ifthenelse{\boolean{articletitles}}{{\it {Measurement of the $B^+$, $B^0$,
  $B_s^0$ meson and $\Lambda_b^0$ baryon lifetimes, lifetime ratios and
  $\Delta\Gamma_d/\Gamma_d$}}, }{} {LHCb-PAPER-2013-065}, {in
  preparation}\relax
\mciteBstWouldAddEndPuncttrue
\mciteSetBstMidEndSepPunct{\mcitedefaultmidpunct}
{\mcitedefaultendpunct}{\mcitedefaultseppunct}\relax
\EndOfBibitem
\end{mcitethebibliography}
